\documentclass{article}
\usepackage{graphicx} 
\usepackage{abstract} 
\usepackage{amsfonts,amsmath,amssymb,mathtools}
\usepackage[english]{babel}
\usepackage{caption}
\usepackage{pstricks}
\usepackage{subfig}
\onecolumn
\usepackage[utf8]{inputenc}
\usepackage[left=2cm, right=2cm, top=2cm, bottom=2cm]{geometry}
\newenvironment{keywords}{\noindent\textbf{Keywords:} }{}
\usepackage{amsmath,empheq}
\usepackage{amsfonts}
\usepackage{amssymb}
\usepackage{nicematrix}
\usepackage{graphicx}
\usepackage{url}
\usepackage{caption}
\usepackage{keyval}
\usepackage{subfig}
\usepackage{placeins}
\usepackage{authblk}
\usepackage{algorithm}
\usepackage{algpseudocode}

\usepackage{amsthm}

\usepackage{xcolor}
\definecolor{verde}{rgb}{0,0.6,0}
\definecolor{darkgreen}{rgb}{0,0.3,0}

\newtheorem{remark}{Remark}
\newtheorem{proposition}{Proposition}
\newtheorem{corollary}{Corollary}

\newtheorem{theorem}{Theorem}

\newtheorem{algo}{Algorithm}

\newtheorem{example}{Example}

\DeclareMathOperator {\diag}{diag}

\begin{document}

\title{Maximum entropy reconstruction of discrete-time Markov chains under partial information} 

\author{Paolo Bolzern, Patrizio Colaneri, Giuseppe De Nicolao}

\date{October 2026}

\maketitle

\begin{abstract}
    This paper addresses the reconstruction of discrete-time Markov chains from incomplete information using the maximum entropy rate principle. Three settings are considered: prescribed graph topology alone, topology with a fully specified stationary distribution, and topology with partially specified stationary probabilities. We investigate the relationship between entropy-maximizing transition mechanisms and Luce's choice axiom, establishing conditions under which optimal transition matrices admit a Luce-score representation. When the stationary distribution is fully prescribed, we develop a constructive feasibility test and a system-theoretic iterative algorithm based on a Lyapunov-like function. Unlike conventional matrix-scaling approaches, the proposed algorithm operates directly on Luce scores. We illustrate the results for selected graph topologies, including complete graphs without self-loops and odd cycles. We also provide a counterexample to a previously stated uniqueness claim, showing that distinct, nonproportional Luce-score vectors can generate the same entropy-maximizing transition matrix. For partially prescribed stationary distributions, we extend the score-based approach to jointly reconstruct the unknown stationary probabilities and transition mechanism through an iterative procedure involving a parameterized Lyapunov-like function. Overall, the proposed framework integrates graph compatibility, entropy-rate maximization, and probabilistic choice representations, offering constructive tools for reconstructing Markov chains under different levels of stationary information and structural constraints.
\end{abstract}


\section{Introduction}

Markov chains provide fundamental models of stochastic dynamics on discrete state spaces, with applications ranging from statistical physics and network science to web ranking, biological processes, and control systems \cite{Kemeny69,Pastor15,Proskurnikov17,BCDAutomatica19,Kumar15}. In many applications, however, transition probabilities are not directly available. The information at hand may be limited to the underlying graph, a prescribed stationary distribution, or only selected components of that distribution. Reconstructing a discrete-time Markov chain from such incomplete information is therefore a natural inverse problem, generally admitting multiple compatible transition mechanisms.

Related reconstruction problems arise in several areas. In ranking and network analysis, one seeks transition mechanisms consistent with observed long-run behavior \cite{Kumar15,Mascherpa26}. In probabilistic choice theory, Luce's axiom provides a structured representation of choice probabilities \cite{Luce59}. In statistical physics and graph dynamics, the maximum entropy random walk (MERW) selects a specific transition mechanism from graph topology by maximizing the entropy rate \cite{Parry64,Burda09,Delvenne11,Duda12}. Related entropy-based models have also been investigated in robotic surveillance \cite{Bullo19}. The presence of forbidden transitions connects reconstruction to transportation problems, network-flow feasibility, and matrix scaling \cite{Bolker72,Gale57,Sinkhorn67,Franklin89}.

In this paper, we distinguish three settings: the graph topology alone is known; the topology and the full stationary distribution are prescribed; or the topology is known together with only partial information on the stationary distribution. In the first setting, the graph determines the admissible transitions but generally does not identify a unique chain. In the other two settings, compatibility between the graph and the stationary constraints must also be considered. Across all the three settings, a further question is how to select one transition mechanism among those compatible with the available information. Maximization of the entropy rate provides a common selection principle \cite{Jaynes57,Justesen84,Girardin04}. An alternative perspective, particularly relevant when the stationary distribution is specified, is provided by reconstruction methods based on Luce's choice axiom \cite{Kumar15,Luce59}.

In the topology-only problem, Parry used the Perron–Frobenius eigenvectors of the adjacency matrix to define the transition probabilities that maximize entropy rate \cite{Parry64}. The subsequent MERW literature highlighted distinctive properties of maximum-entropy walks, including the statistical structure of paths and localization phenomena on irregular graphs \cite{Delvenne11,Burda09,Duda12}.

When the full stationary distribution is prescribed, feasibility becomes a central issue. 
The existence of compatible transition probabilities can be studied in terms of stationary probability flows between states. The requirement that the total incoming and outgoing flows at each state equal its prescribed stationary probability leads to feasibility conditions that can be characterized using classical results from graph theory and network-flow analysis \cite{Hall35,Gale57,Bolker72,Brualdi68}.
Once compatibility has been established, a specific solution can be searched for by entropy-rate maximization, which involves computations closely related to matrix scaling and iterative proportional fitting \cite{Sinkhorn67,Franklin89}. A complementary approach to the reconstruction of transition mechanisms is based on Luce's choice principle \cite{Kumar15,Luce59}. These connections motivate a closer examination of the relationship between entropy-maximizing chains and transition rules parameterized by Luce scores.

The reconstruction of Markov chains from partially prescribed stationary probabilities is related to several established lines of research. Previous contributions have addressed the estimation of transition probabilities from aggregate observations \cite{Kalbfleisch83}, including recent formulations based on inverse optimal transport \cite{Mascherpa26}. These studies provide important theoretical and methodological foundations, but they address different aspects of the reconstruction problem. In the setting considered here, the graph topology is prescribed, while only some stationary probabilities are known. This requires determining the remaining stationary probabilities jointly with the transition mechanism, while ensuring compatibility with the graph. Our work addresses these interconnected requirements through entropy-rate maximization and develops a constructive approach based on Luce scores.

In previous work, the authors of the present paper studied maximum-entropy reconstruction for continuous-time homogeneous Markov chains, considering unconstrained, fully prescribed, and partially prescribed stationary distributions \cite{BCD_Automatica25}. The discrete-time setting raises additional issues. Besides differences in the formulation of the entropy criterion, the prescribed graph may prevent the existence of a chain compatible with the stationary constraints, even when the graph is irreducible and aperiodic. Allowing self-loops can simplify feasibility, as illustrated by related reconstruction settings \cite{Bullo19}, but such an assumption is not imposed here. These differences motivate a separate treatment of discrete-time chains.

Our contributions concern the relationship between entropy-maximizing transition mechanisms, choice axioms, and constructive algorithms. First, we establish conditions under which entropy-maximizing Markov chains admit a Luce-type representation. Second, for a fully prescribed stationary distribution, we develop a constructive admissibility test and a system-theoretic algorithm based on a Lyapunov-like function. The algorithm computes entropy-maximizing transition probabilities through Luce scores rather than stationary edge flows, providing an alternative to flow-based and Sinkhorn-type constructions. Our admissibility test also offers an alternative to the combinatorial characterization provided by generalized Hall-type conditions \cite{Gale57}. We illustrate the approach on selected graph topologies, including complete graphs without self-loops and odd cycles. We further revisit a uniqueness claim concerning Luce scores \cite{Kumar15} and present a counterexample in which distinct, non-proportional score vectors induce the same maximum-entropy transition matrix. This distinction between the uniqueness of a transition mechanism and that of its score representation is particularly relevant when scores are used as web ranking indices. Third, we extend the score-based reconstruction approach to partially prescribed stationary distributions, jointly determining the unknown stationary probabilities and the transition mechanism through an iterative algorithm based on a parameterized Lyapunov-like function. Together, these results provide a unified perspective on graph constraints, entropy-rate maximization, and probabilistic choice representations in discrete-time Markov-chain reconstruction.

The remainder of the paper is organized as follows. Section 2 introduces the necessary preliminaries. Section 3 investigates the relationship between entropy-rate maximization and Luce's choice axiom. Section 4 addresses the reconstruction of maximum-entropy Markov chains under three different assumptions on the stationary distribution: completely free, fully prescribed, and partially prescribed. Section 5 examines particular graph topologies, including complete graphs without self-loops, string graphs with a single self-loop, and odd ring graphs without self-loops.  Finally, the Appendix presents a counterexample illustrating the nonuniqueness of the normalized Luce-score vector.

\vspace{0.5cm} 
\noindent
\emph{Notation.} For a nonnegative matrix $A$, the symbol  $\lambda_F(A)$ denotes the Perrin-Frobenius eigenvalue of $A$.

\section{Preliminaries}
Let $\sigma(k) \in {\mathcal M}=\{1,2,\ldots,M\}$ be a discrete-time homogeneous Markov chain with transition probability matrix $P\in\mathbb{R}^{M \times M}$, where $p_{ij}={\rm Prob}(\sigma(k+1)=j|\sigma(k)=i)$. The set of row-stochastic matrices $P$ of size $M$ will be denoted by $\mathcal{S}$.
Note that $P{\bf 1}_M={\bf 1}_M$, where ${\bf 1}$ is a vector of ones of dimension $M$. The chain topology is described by the adjacency matrix $\Gamma$ such that $\Gamma_{ij}=1$, if $p_{ij}>0$, and $\Gamma_{ij}=0$ otherwise, i.e. there is no link from node $i$ to node $j$. If $\Gamma=\Gamma'$ the graph is said to be undirected, otherwise it is directed. The graph is said to have the $i$-th self-loop whenever $\Gamma_{ii}=1$.
For a given subset $\mathcal A \subset \mathcal M$, the sets $\Gamma_{in}(\mathcal A)$ and $\Gamma_{out}(\mathcal A)$ will denote the sets of in-neighbors and out-neighbors nodes of set $\mathcal A$, respectively.
The symbol $\mathcal{S}(\Gamma)$ will denote the subset of matrices $P \in \mathcal{S}$ consistent with $\Gamma$, i.e. $p_{ij}=0$ if and only if $\Gamma_{ij}=0$.

In the following, it is assumed that $\Gamma$ is irreducible and aperiodic \cite{Cassandras08}.
Recall that irreducibility implies that every state in the graph is reachable from every other state, while aperiodicity ensures that the return time to any state is not constrained to a multiple of an integer greater than $1$.
Under these assumptions, the process $\sigma(k)$ is ergodic and there exists a unique steady-state probability distribution $\pi$, belonging to the open simplex $\Delta$ of strictly positive unit-sum vectors, satisfying
{\begin{equation}
\label{eq.statprob}
  P^\top \pi = \pi  
\end{equation}}

According to \cite{Shannon48}, the entropy (rate) of a discrete-time Markov chain is given by
\begin{equation}
\label{eq.H}
H=-\sum_i\pi_i \sum_{j}p_{ij}{\rm log}(p_{ij})
\end{equation}
Above, it is tacitly assumed that $p_{ij}\log(p_{ij})=0$ whenever $p_{ij}=0$.

If $\pi$ is fixed, it is immediate to see that $H$ is a strictly concave function of the transition probabilities $p_{ij}$. As discussed later, a consequence is the uniqueness of the maxentropic chain consistent with a given steady-state probability distribution $\pi$. 

The chain with transition probability matrix $P$ is said to fulfill the Luce's choice axiom \cite{Luce59}, if there exists a strictly positive vector $s \in \mathbb R^M$ of dimension $M$ such that
\begin{eqnarray}
\label{eq.Luce}
    p_{ij}=\frac{\Gamma_{ij} s_j}{\sum_k \Gamma_{ik} s_k}.
\end{eqnarray}
Given a pool of alternative items, the axiom says that the relative odds of choosing one item over another do not depend on the presence or absence of other items in the pool.
In other words, each item has a latent score $s_i$ and the items are chosen based on their score. The axiom is also associated with the so-called principle of independence of irrelevant alternatives (i.i.a.). Of course, the set of Markov chains that fulfill Luce's axiom is parameterized by the $M$-dimensional positive score vector $s$.

In the paper, the notation $x<y$ or $x\le y$, where $x$ and $y$ are real vectors, stands for $x_i < y_i$ or $x_i \le y_i, \forall i$, respectively. 

In a directed graph, the symbols $\Gamma_{in}(\mathcal A)$ and $\Gamma_{out}(\mathcal A)$ represent the subset of in-neighbors and out-neighbors of a set $\mathcal A$,  respectively.  

\section{Maximization of entropy and Luce's choice axiom}

For a given $\pi \in \Delta$,  consider the problem of maximizing $H$ with respect to $p_{ij}$. For this purpose, we introduce the Lagrangian
$$
\mathcal L(P) = -\sum_i\pi_i \sum_{j}p_{ij}{\rm log}(p_{ij}) + \lambda^\top(P^\top \pi-\pi) + \mu^\top(P{\bf 1}_M-{\bf 1}_M)
$$
where $\lambda$ and $\mu$ are multiplier vectors of dimension $M$. In order to derive necessary conditions for entropy maximization, stationarity of the Lagrangian is imposed. Differentiating with respect to $p_{ij}$, we have
$$
\frac{\partial \mathcal L(P)}{\partial p_{ij}} = - \pi_i \left(\log(p_{ij}) + 1  \right) + \lambda_j \pi_i + \mu_i.
$$
 By zeroing the derivatives and recalling that $p_{ij}=0$ if $\Gamma_{ij}=0$, it follows that
$$
p_{ij} = \Gamma_{ij} e^{(\lambda_j -1 )} e^{(\mu_i/\pi_i )}. 
$$
Defining $x_i=e^{(\lambda_i -1 )}$ and $y_i= \sum_j \Gamma_{ij}x_j$, from $\sum_j p_{ij}=1, \forall i$, it follows that $e^{(\mu_i/\pi_i )}=1/y_i$, so that
  
\begin{equation}
\label{eq.para}
p_{ij}=\Gamma_{ij}\frac{x_j}{y_i}, \quad y=\Gamma x.
\end{equation}
Finally, the congruence equation $P^\top \pi=\pi$ can be written as 
\begin{equation}
\label{eq.cong}
x_j \sum_i \pi_i\Gamma_{ij}\frac{1}{y_i}=\pi_j, \quad \forall j.
\end{equation}
Letting $X={\rm diag}(x)$ and $Y={\rm diag}(y)={\rm diag}(\Gamma x)$, equations \eqref{eq.para} and \eqref{eq.cong} can be equivalently rewritten as
\begin{equation}
\label{eq.paramat}
P=Y^{-1}\Gamma X
\end{equation}
\begin{equation}
\label{eq.congmat}
\pi^\top Y^{-1}\Gamma X=\pi^\top.
\end{equation}
It is worth noting that, due to the strict concavity of $H(P)$, and the linearity of the constraints, for a given $\pi \in \Delta$, any stationarity point of the Lagrangian identifies a maxentropic chain. Moreover, thanks to strict concavity, such a maximizing solution is also unique.

Note that the $M$-dimensional vectors $x$ and $y$ are strictly positive, so that $p_{ij}>0$ whenever $\Gamma_{ij}>0$. Therefore, $P$ defines an irreducible and aperiodic chain, implying that \eqref{eq.congmat} admits a unique solution $\pi \in \Delta$ for any $x$ in the positive orthant. In the sequel, $\pi=\phi(x)$ denotes the function that returns the solution $\pi$ of \eqref{eq.congmat} for any given vector $x>0$.
It is easily seen that 
\begin{equation}
\label{eq.phi}
\pi=\phi(x)=(X \Gamma^\top Y^{-1} + {\bf 1}{\bf 1}^\top - I)^{-1} {\bf 1}\quad ,\quad X={\rm diag}(x)\quad,\quad Y={\rm diag}(\Gamma x).
\end{equation}
The function $\phi(x)$ is the extension to the case of nonsymmetric graph matrices $\Gamma$ of the maxentropic vector map introduced and studied by \cite{Bullo19} under the assumption $\Gamma=\Gamma^\top$, i.e. for directed Markov chains. In the undirected case, the expression of $\phi(x)$ can be greatly simplified as shown in the following proposition. In \cite{Bullo19},  the property was proven for graphs with all self-loops and it was conjectured that it was true also in the general case. Here, we prove the conjecture.

\begin{proposition}
\label{pro.symmetric}
If the Markov chain graph is undirected, i.e. $\Gamma=\Gamma^\top$, then
\begin{equation}
\label{eq.phi_sym}
\pi=\phi(x)= \frac{X \Gamma x}{x^\top \Gamma x}=
\frac{x \odot y}{x^\top y}
\end{equation}
\end{proposition}
\begin{proof}
Since $\Gamma=\Gamma^\top$,
$$
\phi(x)=(X \Gamma Y^{-1} + {\bf 1}{\bf 1}^\top - I)^{-1} {\bf 1}.
$$
Now, compute the expression
\begin{eqnarray*}
&&(X \Gamma Y^{-1} + {\bf 1}{\bf 1}^\top - I)X\Gamma X {\bf 1}\\
&=&X \Gamma X Y^{-1} \Gamma X {\bf 1} + {\bf 1}{\bf 1}^\top X \Gamma X {\bf 1} - X\Gamma X {\bf 1}.
\end{eqnarray*}
The above computation takes advantage of the fact that the diagonal matrices $Y^{-1}$ and $X$ commute. Since
$$
Y^{-1} \Gamma X {\bf 1}={\bf 1}
$$
by the very definition of $Y$, it follows that
$$
(X \Gamma Y^{-1} + {\bf 1}{\bf 1}^\top - I)X\Gamma X {\bf 1} = {\bf 1} ({\bf 1}^\top X \Gamma X {\bf 1}).
$$
Hence,
$$
(X \Gamma Y^{-1} + {\bf 1}{\bf 1}^\top - I)^{-1} {\bf 1}=\frac{X \Gamma X {\bf 1}}{{\bf 1}^\top X \Gamma X {\bf 1}}
$$
so that the thesis follows directly.
\end{proof}

Note that the knowledge of $x$ is sufficient to obtain $y$, through $y=\Gamma x$, and $P$, through \eqref{eq.para}, so that the maxentropic Markov chain is completely specified, including its steady-state probability distribution $\pi$. Notice that applying a scaling factor to $x$ does not affect the computation of $P$, so that $\phi(\alpha x)=\phi(x)$, for any positive scalar $\alpha$. Hence, the set of maxentropic Markov chains can be parameterized in several ways, e.g. by assuming that $x$ belongs to the open simplex $\Delta$, i.e. $\sum_i x_i =1, x_i >0, \forall i$,  or by fixing one of the entries of $x$. Observe that, in general, it is not possible to parametrize the same set using the steady-state probability distribution $\pi$, because the congruence equation \eqref{eq.cong} may not be solvable in the unknown $x$. As shown in \cite{Bullo19} for undirected graphs, an important exception is the case when all self-loops are present, namely $\Gamma_{ii}=1, \forall i$.

Quite remarkably, by comparing \eqref{eq.para} with \eqref{eq.Luce}, it turns out that $x$ can be interpreted as a score vector according to Luce's choice axiom. Therefore, we have the following proposition.

\begin{proposition}
A maxentropic chain is characterized by the positive real-valued vector $x$ of dimension $M$ and satisfies Luce's choice axiom with score vector $s=x$. 
\end{proposition}

It is interesting to discuss  *** see web ranking / graph centrality *** the issue of uniqueness of the normalized (e.g. in the open simplex) score vector $x$ associated with a given maxentropic chain.
This question was addressed in Theorem 11 of \cite{Kumar15}, claiming the uniqueness of the normalized score vector. Unfortunately, the result is false as shown by the following counterexample.

\begin{example}
\label{ex.nonunique}
Consider the irreducible aperiodic Markov chain with graph adjacency matrix
$$
\Gamma=\begin{bmatrix}
0 & 1 & 1\\
1 & 0 & 0\\
0 & 1 & 1    
\end{bmatrix} 
$$

It is easy to verify that both score vectors (lying in the open simplex $\Delta$)
\begin{eqnarray*}
    x_a &=& \begin{bmatrix}
        1/3 & 1/3 & 1/3
    \end{bmatrix}^\top\\
    x_b &=& \begin{bmatrix}
        1/2 & 1/4 & 1/4
    \end{bmatrix}^\top
\end{eqnarray*}
generate, from \eqref{eq.paramat}, the same transition probability matrix
$$
P = \begin{bmatrix}
0 & 0.5 & 0.5\\
1 & 0 & 0\\
0 & 0.5 & 0.5
\end{bmatrix}
$$
with associated steady-state probability distribution $\pi=\frac{1}{3}\,{\bf 1}$.  Hence, the normalized score of the maxentropic chain identified by $P$ is not unique.

In order to explain such a circumstance, note that the in-neighbor set of the state subset $\mathcal A=\{1\}$ is given by $\mathcal B=\{2\}$, with $\mathcal A$ being the out-neighbor set of $\mathcal B$. In view of this, the score of state $1$ can be scaled independently of the scaling applied to the remaining states, without affecting the transition probability matrix P. In fact, $x_b$ is obtained from $x_a$ by doubling the first entry and renormalizing the entire vector.
\end{example}

As shown in Example \ref{ex.nonunique}, the situation of nonunique normalized scores arises whenever there exists a proper state subset $\mathcal A$ such that $\mathcal A=\Gamma_{out}(\Gamma_{in}(\mathcal A))$. The issue of the nonuniqueness of the normalized scores is treated in further detail in the Appendix, where, among other things, it is shown that uniqueness is guaranteed in the case of (irreducible and aperiodic) undirected chain graphs. Concerning the maxentropic vector map $\phi(x): \Delta \mapsto \Delta$, we can therefore say that it is surjective but not injective, with the exception of undirected chains for which it is bijective.

Next, we show that the maximal value of the entropy can be expressed as a simple function of the score vector $x$. 
*** citare Sinkhorn ***

\begin{proposition}
\label{re.entropy_x}

For a maxentropic Markov chain with adjacency matrix $\Gamma$ the entropy is given by
\begin{equation}
H(x)= \sum_i\pi_i{\rm log}\left(\frac{y_i}{x_i}\right)
\label{eq.entr}
\end{equation}
where $y=\Gamma x$, and $\pi$ satisfies \eqref{eq.cong}. Moreover, when $\Gamma_{ii}=1, \forall i$, i.e. the graph possesses all self-loops, then $H(x)=-\sum_i \pi_i \log(p_{ii}) $.
\end{proposition}

\begin{proof}
For a given vector $x>0$, we have
$$
H(x)=-\sum_i\pi_i\sum_j \Gamma_{ij}\frac{x_j}{y_i}\left({\rm log}(x_j)-{\rm log}(y_i)\right)=
-\sum_i\pi_i\sum_j \Gamma_{ij}\frac{x_j}{y_i}{\rm log}(x_j)+\sum_i\pi_i{\rm log}(y_i)
$$
In view of the congruence equation \eqref{eq.cong}, it turns out that

\begin{eqnarray*}
H(x)&=&-\sum_j\left(\sum_i\pi_i 
\Gamma_{ij}\frac{1}{y_i}x_j\right){\rm log}(x_j)+\sum_i\pi_i{\rm log}(y_i) \nonumber \\
&=& \sum_i\pi_i{\rm log}\left(\frac{y_i}{x_i}\right)
\end{eqnarray*}
In the case $\Gamma_{ii}=1, \forall i$, it suffices to recall \eqref{eq.para}.
\end{proof}

\section{Maxentropic chains}
In this section, we study the problem of finding the maxentropic chain  considering free, constrained, or partially constrained steady-state distribution.
\subsection{Free stationary distribution}
Assume that the steady-state probability distribution $\pi$ is completely free. Entropy maximization under this scenario was thoroughly investigated in \cite{Duda12}. Here, we revisit the same optimization problem by exploiting the parametrization results of the previous section.

Consider the maximization of $H$ in \eqref{eq.entr} with respect to $\pi$ under the constraint $\sum_i\pi_i=1$. Letting 
${\cal L}=H+\eta(-1+\sum_i\pi_i)$ and zeroing the derivatives of ${\cal L}$ with respect to $\pi_i$, it follows that $\eta={\rm log}(\frac{y_i}{x_i})$ for all $i$, i.e. $y_i=\alpha x_i$, $\alpha=e^{\eta}$. Therefore, \eqref{eq.entr} implies
$H=\log(\alpha)$. 
In view of the relationship $y=\Gamma x$, we have that 
$\Gamma x=\alpha x$. Recalling that $\Gamma$ is a nonnegative matrix and all the entries in $x$ are strictly positive, we see that $\alpha$ coincides with $\lambda_F$, the Perron-Frobenius eigenvalue, and the score vector $x$ with the  Perron-Frobenius right eigenvector  of $\Gamma$.
From the congruence equation \eqref{eq.congmat}, we also have that 
$$
\pi^\top X^{-1}\Gamma=\lambda_F\pi^\top X^{-1}
$$
so that $X^{-1}\pi$ is the left Perron-Frobenius eigenvector of $\Gamma$. This allows to formulate the solution of the entropy maximization problem in terms of the Perron-Frobenius eigenvectors of the adjacency matrix.

More precisely, letting $w$ and $v$ be any right and left Perron-Frobenius eigenvectors of $\Gamma$, i.e. $\Gamma w=\lambda_F w$ and $\Gamma^\top v=\lambda_F v$, the maxentropic solution for the free stationary distribution case is given by
\begin{equation}
\label{eq.pifree}
\pi^*=\frac{v\odot w}{v^\top w}, \quad p^*_{ij}=\Gamma_{ij}\frac{w_j}{\sum_k\Gamma_{ik}w_k}=\Gamma_{ij}\frac{w_j}{\lambda_F w_i}, \quad H^*={\rm log}(\lambda_F)
\end{equation}

In fact, \eqref{eq.pifree} coincides with the result derived in \cite{Duda12} on the MERW (Maxentropic Random Walk). The new insight brought about by our analysis concerns the interpretation of the right eigenvector $w$ in terms of the optimal choice of the Luce's scores $x$. Indeed, for a given adjacency matrix $\Gamma$, i.e. a fixed graph topology, there exist infinitely many score vectors $s$ generating transition probabilities that fulfill Luce's axiom. Entropy maximization provides a rational basis for the choice of a unique chain characterized by the transition probability matrix $P^*$.

Remarkably, the left eigenvector $v$ provides the maxentropic scores of the reversed Markov chain, obtained by replacing $\Gamma$ with $\Gamma^\top$. Finally, it is interesting to note that, for undirected graphs, the maxentropic steady-state probabilities $\pi_i^*$ are proportional to the squared scores $x_i^2$. To the authors' knowledge, this connection between Luce's scores and maxentropic steady-state probabilities was not highlighted before.

\begin{remark}
\label{re.centrality}
Entropy maximization in the case of the free stationary distribution can be regarded as a method to derive a centrality score for the nodes of the graph with adjacency matrix $\Gamma$. A classic centrality measure [Bonacich 1972] is the eigenvector centrality score, namely the right eigenvector $w$ of \eqref{eq.pifree}. 
Notable the celebrated PageRank score is strictly related to the eigencentratlity score of the reversed graph, namely the left eigenvector $v$ of \eqref{eq.pifree}.
As an alternative to these scores, one may resort to the maxentropic centrality score vector defined by $\pi^*$ in \eqref{eq.pifree}, whose entries are proportional to the products $w_i v_i$. The maxentropic centrality score of a node is the product of its eigencentrality $w_i$ for the graph $\Gamma$, and its eigencentrality $v_i$ for the reversed graph $\Gamma^\top$. If the graph is undirected, i.e. $\Gamma=\Gamma^\top$, then $\pi^*\propto w \odot w$, so that classical eigencentrality and maxentropic centrality yield the same ranking of nodes. 
\end{remark}

\subsection{Fixed stationary distribution}

When the adjacency matrix $\Gamma$ and the stationary distribution $\pi$ are given, before proceeding with the calculation of the maxentropic $P$, one has to answer a preliminary question about the existence of a transition probability matrix $P$ such that \eqref{eq.statprob} is satisfied.

\subsubsection{Consistency of the stationary distribution}
For arbitrary $\Gamma$ and a given $\pi$, eq. \eqref{eq.statprob} may not admit any solution $P\in \mathcal{S}(\Gamma)$. 
In the following, $\mathcal{D}(\Gamma)$ will denote the set of stationary distributions $\pi \in \Delta$ such that \eqref{eq.statprob} is satisfied for some $P \in \mathcal{S}(\Gamma)$:
$$
\mathcal{D}(\Gamma)=\{\pi\in \Delta: \exists P \in \mathcal{S}(\Gamma), \pi^\top P = \pi^\top\}.
$$
When $\pi \in \mathcal{D}(\Gamma)$, we will say that $\pi$ is \emph{consistent} with $\Gamma$. The set of associated transition probability matrices $P$ is defined as
$$
\mathcal{P}(\Gamma, \pi)=
\left \{ P \in \mathcal{S}(\Gamma):\pi^\top P= \pi^\top, \pi \in \Delta
\right \}.
$$
It is immediate to see that $\mathcal{P}(\Gamma, \pi)$ is empty if and only if $\mathcal{D}(\Gamma)$ is empty.

\begin{example}
\label{ex.consist1}
Consider a 3-state Markov chain with adjacency matrix
$$
\Gamma = \begin{bmatrix}1 & 1 & 0 \\1 & 0 & 1 \\0 & 1 & 0\end{bmatrix}
$$
It is easy to see that 
$$
\mathcal{D}(\Gamma)=\left \{ \pi\in \Delta: \pi_2 > \pi_3, \pi_2 < 0.5\right \}.
$$
Moreover, for any $\pi \in \mathcal{D}(\Gamma)$, the (unique)  transition probability matrix $P \in \mathcal{P}(\Gamma, \pi)$ is
$$
P = \left[\begin{array}{ccc} \pi_1 & 0 & 0 \\0 & \pi_2 & 0 \\0 & 0 & \pi_3\end{array}\right]^{-1}
\left[\begin{array}{ccc}1-2\pi_2 & \pi_2-\pi_3 & 0 \\\pi_2-\pi_3  & 0 & \pi_3 \\0 & \pi_3 & 0\end{array}\right].
$$
Note that, in this example, the cardinality of $\mathcal{P}(\Gamma, \pi)$ is one. 
\end{example}

\begin{example}
\label{ex.consist2}
This example shows that there might be infinitely many $P \in \mathcal{P}(\Gamma, \pi)$ associated with a single $\pi \in \mathcal{D}(\Gamma)$. For instance, consider
$$
\Gamma = \begin{bmatrix} 1 & 1 & 0 \\1 & 1 & 1 \\0 & 1 & 0\end{bmatrix}.
$$
It is immediate to see that
$$
\mathcal{D}(\Gamma)=\left \{ \pi\in \Delta: \pi_2 > \pi_3, \pi_3 < 0.5\right \}.
$$
In this case, the cardinality of $\mathcal{P}(\Gamma, \pi)$ is actually infinite $\forall \pi \in \mathcal{D}(\Gamma)$. Precisely for any given $\pi \in \mathcal{D}(\Gamma)$, the set $\mathcal{P}(\Gamma, \pi)$ is parameterized as follows

$$
P = 
\begin{bmatrix}
p_{11} & 1-p_{11} & 0 
\\ (1-p_{11})\pi_1/\pi_2 & (2 \pi_2 -1 + \pi_1 p_{11})/\pi_2 & \pi_3/\pi_2 
\\ 0 & 1 & 0
\end{bmatrix}
$$
where $p_{11} \in \left(\frac{1-2\pi_2}{\pi_1},1 \right)$.
\end{example}

The previous examples show that, for a given stationary distribution $\pi$, three situations may arise depending on the chain topology $\Gamma$: (i) if $\pi \notin \mathcal{D}(\Gamma)$ the distribution is not consistent and no solution exists; (ii)  the set $\mathcal{P}(\Gamma, \pi)$ has cardinality one so that the solution is unique; (iii) the cardinality of the set $\mathcal{P}(\Gamma, \pi)$ is greater than one (actually infinite) and it is meaningful to look for a solution that maximizes the entropy.

The complete characterization of $\mathcal{D}(\Gamma)$ is an interesting problem that can be reformulated as a feasible flow problem in a graph with forbidden edges. In the relevant literature, the necessary and sufficient condition for consistency is a weighted version of Hall's marriage theorem that involves a combinatorial set of linear inequalities on the steady-state probability distribution, see \cite{Brualdi68}. Notably, the criterion is polynomial-time checkable because, rather than checking all node subsets, one can solve a single maximum-flow problem.

Before concluding the subsection, we provide a simple sufficient condition for consistency in a special case and a necessary condition of general validity. If the chain graph possesses all self-loops, i.e. $\Gamma_{ii}=1, \forall i$, then $\mathcal{D}(\Gamma)$ is nonvoid and coincides with the open simplex $\Delta$, see \cite{Bullo19} for the case of undirected graph,  $\Gamma=\Gamma^\top$, and \cite{Kumar15} for the general case. Finally, an easy-to-check necessary condition is the following.

\begin{proposition}
\label{cor.onehalf}
A necessary condition for the consistency of $\pi$, i.e. $\pi \in \mathcal{D}(\Gamma)$, is \begin{equation}
\label{eq.pi_consist}
    \pi^\top \le \pi^\top \Gamma
\end{equation}
Moreover, if $\Gamma_{ii}=0$ for some $i$, i.e. the $i$-th state does not have a self-loop, $\pi$ is consistent, i.e. $\pi \in \mathcal{D}(\Gamma)$, only if $\pi_i < 0.5$, i.e.
$$
\Gamma_{ii}=0 \longrightarrow \pi_i < 0.5
$$ 
\end{proposition}
\begin{proof}
Since $P \le \Gamma$, we have that
$
    \pi^\top = \pi^\top P \le \pi^\top \Gamma.
$
Moreover, from \eqref{eq.pi_consist}, we see that
 $$
 \pi_i \le \sum_{k \ne i} \pi_k \Gamma_{ki} + \pi_i \Gamma_{ii} = \sum_{k \ne i} \pi_k \Gamma_{ki} \le \sum_{k \ne i} \pi_k = 1-\pi_i
$$
so that  $\pi_i \le 0.5$. Next, we prove that this inequality is actually strict. Assume by contradiction that $\pi_i=0.5$. Since
$$
0.5=\pi_i= \sum_{j \ne i}\Gamma_{ji}p_{ji} \pi_j
$$
and 
$
\sum_{j \ne i}\pi_j = 0.5
$
it follows that $p_{ji}=\Gamma_{ji}=1, j \ne i$.
This corresponds to a star-topology with the $i$-th state as center and without self-loops, thus violating the aperiodicity assumption. 
\end{proof}
\subsubsection{Computation of the maxentropic chain}
Turning now to the problem of finding the maxentropic chain for a fixed $\pi \in \Delta$, we consider the parameterization of maxentropic chains provided by \eqref{eq.paramat}, \eqref{eq.congmat} and \eqref{eq.phi}. 
If $\pi \in \mathcal{D}(\Gamma)$, then $\mathcal{P}(\Gamma, \pi)$ is nonempty and there exists an associated maxentropic chain. Recalling \eqref{eq.phi}, we can use the score vectors $x$ to parametrize the set of consistent steady-state distributions:
\begin{equation}
\label{eq.x_param}
 \mathcal{D}(\Gamma)=\{\pi: \pi=\phi(x), x \in \Delta\}.
\end{equation}
In conclusion, searching for the maxentropic chain for a fixed $\pi \in \mathcal D(\Gamma)$ amounts to inverting the function $\pi=\phi(x)$, thus obtaining the score vector $x=\phi^{-1} (\pi)$. Once $x$ is obtained, the maxentropic transition probability matrix $P$ and the associated entropy $H$ can be computed from \eqref{eq.paramat} and \eqref{eq.entr}, respectively. In general, the inversion of $\pi=\phi(x)$ is not an easy task due to the nonlinearity of the function $\phi(x)$. In the following, we propose an iterative algorithm that converges to a vector $x$ such that $\pi = \phi(x)$. The algorithm is based on the congruence equation \eqref{eq.congmat}, which can be rewritten in scalar form as
\begin{equation}
\label{eq.congr_scal}
x_i=\frac{\pi_i}{\sum_j \Gamma_{ji} \frac{\pi_j}{y_j}}, \quad \forall i \in \mathcal M
\end{equation}

\medskip
\begin{algo}
\label{algo.1}
\,
\begin{itemize}
    \item {\bf Step 0.}
    $\Gamma$ and $\pi$ are given. 
    \item {\bf Step 1.}
    Let $\varepsilon>0$ be a sufficiently small real scalar and $L>0$ a sufficiently large real scalar. Set $k=0$, $z_i^{[0]}=1, i \in \mathcal M$.
    \item {\bf Step 2.} 
    For each $j \in \mathcal M$, compute
    \begin{equation}
    \label{eq.dyn_sys_z1}
    y_j^{[k]}=\sum_q \Gamma_{jq}z_q^{[k]} .
    \end{equation}
    \item {\bf Step 3.}
    For each $i \in \mathcal M$, compute
    \begin{equation}
    \label{eq.dyn_sys_z2}
    z_i^{[k+1]}=
    \frac{\pi_i}{\sum_j \Gamma_{ji}\frac{\pi_j}{y_j^{[k]}}} .
    \end{equation}
    \item {\bf Step 4.}
    If $\|z^{[k+1]}-z^{[k]}\| < \varepsilon \|z^{[k]}\|$, then define the solution $x=\frac{z^{[k+1]}}{{\bf 1}^\top z^{[k+1]}}$ and quit. Otherwise go to Step 5.
    \item {\bf Step 5.}
    If $\max_i z_i^{[k+1]}/\min_i z_i^{[k+1]} > L$, quit without a solution. Otherwise, set $k \leftarrow k+1$ and go to Step 2.
\end{itemize}
\end{algo}
\medskip

In Algorithm 1, the vector $z^{[k]}$
represents the current value at step $k$ of the unscaled score vector. Note that such a vector remains strictly positive at each iteration. The algorithm terminates either when the relative difference between two consecutive values of $z$ is below the assigned threshold $\epsilon$, or when the ratio between the largest and the smallest entries is above the upper threshold $L$. In the former case, the algorithm is converging to a positive solution $x$ of the congruence condition \eqref{eq.congr_scal} in the open simplex $\Delta$. Once the score vector $x$ has been found, \eqref{eq.para} provides the entries of the transition probability matrix $P$ of the maxentropic chain with the assigned steady-state probability distribution $\pi$. The latter case occurs when the dynamical system defined by \eqref{eq.dyn_sys_z1}, \eqref{eq.dyn_sys_z2} does not admit an equilibrium point and there is no positive finite solution of \eqref{eq.congr_scal}. The convergence result proven in the next theorem implies that, for small enough $\epsilon$ and large enough $L$, the algorithm converges if and only if the stationary distribution $\pi$ is consistent.

\begin{theorem}
The state $z^{[k]}$ of the dynamical system \eqref{eq.dyn_sys_z1}, \eqref{eq.dyn_sys_z2} converges to an equilibrium for any positive initialization $z^{[0]}$ if and only if $\pi \in \mathcal D(\Gamma)$.  
\end{theorem}

\begin{proof}
\emph{Sufficiency}. For a given $\pi \in \mathcal D(\Gamma)$, consider the dynamical discrete-time positive system described in Steps 2 and 3. 
The set of equilibrium points will be denoted by $\mathcal{Z}$.
In view of the parameterization \eqref{eq.x_param} of $\mathcal D(\Gamma)$, the set $\mathcal Z$ is not empty.
Next, we show that the function $V(z)=W(z)-1$ with
$$
W(z)=\max_i \frac{\pi_i}{z_i\sum_j \Gamma_{ji}\frac{\pi_j}{y_j}}
$$
acts as a Lyapunov-like function for the system. 
To this purpose, note that $V(z)$ is nonnegative definite  in the open positive orthant. 
To show this property, assume, by contradiction, that $V(z)<0$, namely $W(z)<1$, for some $z>0$.
Observe that 
$$W(z)= \max_i \frac{\pi_i}{\pi^\top Y^{-1}\Gamma e_i z_i}$$
where $Y={\rm diag}(\Gamma z)$ and $e_i$ is the $i$-th column of the identity matrix. Then, it follows that
$
\pi_i < \pi^\top Y^{-1}\Gamma e_i z_i,\forall i
$.
Summing over all indices $i$ and recalling that $y=\Gamma z$, we get $1 = \sum_i\pi_i <\pi^\top Y^{-1}\Gamma z= \pi^\top{\bf 1}=\sum_i\pi_i=1$ , a contradiction.
Moreover, $V(z)=0$ if and only if $z$ belongs to the set $\mathcal{Z}$ of positive equilibria of the dynamical system \eqref{eq.dyn_sys_z1}, \eqref{eq.dyn_sys_z2}.  Next, we prove that $V(z^{[k+1]})\le V(z^{[k]})$.
Indeed, notice that 
$$
z_i^{[k+1]}\le W(z^{[k]}) z_i^{[k]}, \quad y_j^{[k+1]}\le W(z^{[k]}) y_j^{[k]}
$$
Therefore
\begin{eqnarray*}
W(z^{[k+1]})&=&
\max_i \frac{\pi_i}{z_i^{[k+1]}\sum_j \Gamma_{ji}\frac{\pi_j}{y_j^{[k+1]}}}\\
&\leq&
\max_i \frac{\pi_i}{z_i^{[k+1]}\sum_j \Gamma_{ji}\frac{\pi_j}{W(z^{[k]})y_j^{[k]}}}\\
&=&
\max_i \frac{z_i^{[k+1]} W(z^{[k]})}{z_i^{[k+1]}}= W(z^{[k]})
\end{eqnarray*}
In conclusion,
$V(z^{[k+1]})\le V(z^{[k]})$ and the equal sign holds only when $z^{[k+1]}=z^{[k]}$.
This proves that the set $\mathcal Z$ is globally asymptotically attractive.

\emph{Necessity}. If the state $z^{[k]}$ of the dynamical system \eqref{eq.dyn_sys_z1}, \eqref{eq.dyn_sys_z2} converges to an equilibrium $z$, in view of \eqref{eq.congr_scal}, the associated normalized score vector $x \in \Delta$ is such that $x= \phi^{-1}(\pi)$. Recalling \eqref{eq.x_param}, this implies $\pi \in \mathcal D(\Gamma)$.
\end{proof}

The Lyapunov-like function $V(z)= W(z)-1$ employed in the proof admits an interesting interpretation. Let
$$\pi_i^{[k]}=\sum_j \frac{\Gamma_{ji}\pi_j z_i^{[k]}}{y_j^{[k]}}
$$
be the stationary probabilities associated with the current score vector $z^{[k]}$ or, equivalently, $\pi^{[k]}=\phi(z^{[k]})$. Then, we have that 
$$
W(z^{[k]})=\max_i \frac{\pi_i}{\pi_i^{[k]}} 
$$
Hence, $V$ tends to zero as the current probability distribution $\pi^{[k]}$ approaches the target distribution $\pi$.

\begin{example}
\label{ex.fixed}
Consider a 3-state Markov chain with adjacency matrix
$$
\Gamma = \begin{bmatrix}1 & 0 & 1 \\1 & 1 & 1 \\1 & 1 & 0\end{bmatrix}
$$
If the stationary distribution is completely free, the maxentropic solution is $\pi^*=\begin{bmatrix}0.3493 & 0.3493 & 0.3014 \end{bmatrix}^\top$, with $H^*=0.8096$ and
$$
P^*=\begin{bmatrix}
    0.4450 & 0 & 0.5550 \\
    0.2470 & 0.4450 & 0.3080 \\
    0.3569 & 0.6431 & 0 \\
\end{bmatrix}
$$
Now take $\pi=\begin{bmatrix}0.1 & 0.6 & 0.3 \end{bmatrix}^\top$.
Letting $\varepsilon=10^{-6}$ and $L=10^{6}$, we ran Algorithm \ref{algo.1}. The algorithm converged after $k=11$ iterations, yielding the normalized score vector $x=\begin{bmatrix}
    0.0770 & 0.5604 & 0.3625
\end{bmatrix}^\top$, associated through \eqref{eq.paramat} with the maxentropic transition probability matrix
$$
P=\begin{bmatrix}
    0.1752 & 0 & 0.8248 \\
    0.0770 & 0.5604 & 0.3625 \\
    0.1208 & 0.8792 & 0 \\
\end{bmatrix}
$$
The maximum entropy is $H=0.6909$.
Figure \ref{fig.algo1_pik} shows the behavior of the entries of the current vector $\pi^{[k]} = \phi(z^{[k]})$ during the iterations, clearly demonstrating its convergence towards the target steady-state probability distribution.

\begin{figure}
        \centering
        \includegraphics[width=8cm]{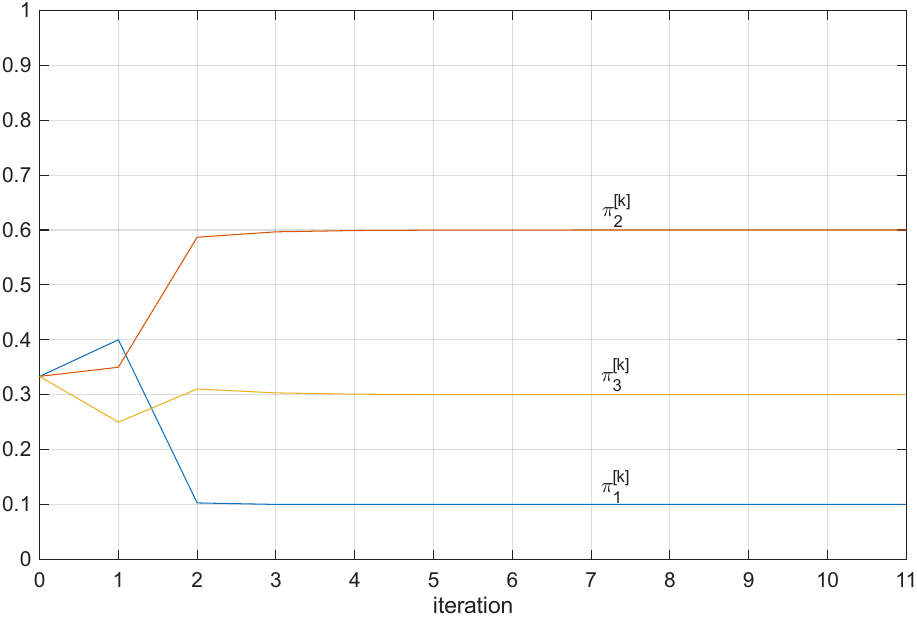}
        \caption{Entries of the current vector $\pi^{[k]} = \phi(z^{[k]})$ during the iterations.}
        \label{fig.algo1_pik}
\end{figure}

Repeating the computation with $\pi=\begin{bmatrix}0.1 & 0.3 & 0.6 \end{bmatrix}^\top$, Algorithm 1 quit after $k=15$ iterations, indicating that such a probability distribution is inconsistent with the given graph topology. Indeed, due to the absence of a self-loop in state $3$, the necessary condition of Theorem \ref{cor.onehalf} is violated.
\end{example}

IDEA: example with non-unique $x$-parametrization of maxentropic chain?

\subsection{Maximization of entropy with partially free stationary distribution}

Given the adjacency matrix $\Gamma$, assume that only some elements of the steady-state probability distribution $\pi$ are known. In this case, the search for the transition probability matrix $P$ of the maxentropic chain involves the determination of the unassigned elements of $\pi$. Our goal is to devise an algorithm to solve this problem.  

Consider the expression \eqref{eq.entr} of the maximum entropy as a function of the positive score vector $x$ and recall that $y=\Gamma x$. Note that \eqref{eq.entr} is valid under the assumption that the congruence equation \eqref{eq.congmat} is fulfilled. Let $\mathcal A \subset \mathcal M$ be the assigned index set, namely the set of indices $i$ for which $\pi_i$ is assigned, and $\mathcal U = \mathcal M  \backslash A$ be the unassigned index set. Denote with $M_a$ and $M_u=M-M_a$ the cardinality of the sets $\mathcal A$ and $\mathcal U$, respectively.
Then, Eq. \eqref{eq.entr} can be decomposed as
$$
H(x) =  \sum_{i \in \mathcal A}\pi_i \log\left(\frac{x_i}{y_i}\right) + \sum_{i \in \mathcal U}\pi_i \log\left(\frac{x_i}{y_i}\right)
$$
We aim to maximize $H$ with respect to $x_i, i \in \mathcal M$, and the elements $\pi_i, i \in \mathcal U$, taking into account the constraint $\sum_{i \in \mathcal M} \pi_i = 1$, the assigned probabilities $\pi_i, i \in \mathcal{A}$. To this purpose, consider the Lagrangian
$$
\mathcal L = H(x) + \eta \left(1 - \sum_{i \in \mathcal A} \pi_i-\sum_{i \in \mathcal U} \pi_i\right)
$$
where $\eta$ is a scalar multiplier. A necessary condition for the optimal solution is that
$$
\frac{\partial \mathcal L}{\partial \pi_i} = \log\left(\frac{x_i}{y_i}\right) - \eta = 0 \quad,\quad i \in \mathcal U
$$
Hence, letting $\beta= \exp{(\eta)}$, it must be 
\begin{equation}
\label{eq.yi_u}
y_i = \beta x_i\quad, \quad i \in \mathcal U
\end{equation}
Now collect the assigned entries of $\pi$ in the vector $\pi_a$, with $\pi_{ai}>0, {\bf 1}_{M_a}^\top \pi_a <1$, collect the unassigned entries in the vector $\pi_u$, and define the permutation matrix $R$ such that
$$
\begin{bmatrix}
    \pi_a \\ \pi_u
\end{bmatrix}
= R \pi = 
\begin{bmatrix}
    R_a \\ R_u
\end{bmatrix}
\pi
$$
Accordingly, define
$$
\begin{bmatrix}
x_a \\ x_u \end{bmatrix}=Rx, 
\quad
\begin{bmatrix}y_a \\ y_u \end{bmatrix}=Ry,
\quad
\begin{bmatrix}
\Gamma_a & \Gamma_{au} \\ \Gamma_{ua} & \Gamma_u
\end{bmatrix}= R \Gamma R^\top
$$
Letting $X_a, X_u, Y_a, Y_u$ denote the positive diagonal matrices associated with the vectors $x_a, x_u, y_a, y_u$, respectively, we have that
$$
X={\rm diag}(x)=R^\top
\begin{bmatrix}
X_a & 0 \\ 0 & X_u
\end{bmatrix}R
\quad,\quad Y={\rm diag}(\Gamma x)=R^\top
\begin{bmatrix}
Y_a & 0 \\ 0 & Y_u
\end{bmatrix}R
$$
Then, the congruence equation \eqref{eq.congmat} becomes
\begin{equation*}
\begin{bmatrix}
\pi_a^\top & \pi_u^\top    
\end{bmatrix}
R =
\begin{bmatrix}
\pi_a^\top & \pi_u^\top    
\end{bmatrix}
RR^\top
\begin{bmatrix}
Y_a^{-1} & 0 \\ 0 & Y_u^{-1}     
\end{bmatrix}
RR^\top
\begin{bmatrix}
\Gamma_a & \Gamma_{au} \\ \Gamma_{ua} & \Gamma_u
\end{bmatrix}
RR^\top
\begin{bmatrix}
X_a & 0 \\ 0 & X_u     
\end{bmatrix}
R
\end{equation*}
or equivalently, since $RR^\top=I_M$,
\begin{equation}
\label{eq.cong_au}
\begin{bmatrix}
\pi_a^\top & \pi_u^\top    
\end{bmatrix}
=
\begin{bmatrix}
\pi_a^\top & \pi_u^\top    
\end{bmatrix}
\begin{bmatrix}
Y_a^{-1} & 0 \\ 0 & Y_u^{-1}     
\end{bmatrix}
\begin{bmatrix}
\Gamma_a & \Gamma_{au} \\ \Gamma_{ua} & \Gamma_u
\end{bmatrix}
\begin{bmatrix}
X_a & 0 \\ 0 & X_u     
\end{bmatrix}
\end{equation}
Considering the block components of the identity \eqref{eq.cong_au}, it results that 
\begin{eqnarray*}
\pi_a^\top &=&
\pi_a^\top Y_a^{-1} \Gamma_{a} X_a
+ \pi_u^\top Y_u^{-1} \Gamma_{ua} X_a\\
\pi_u^\top &=&
\pi_a^\top Y_a^{-1} \Gamma_{au} X_u
+ \pi_u^\top Y_u^{-1} \Gamma_{u} X_u
\end{eqnarray*}
Recalling that $Y_u= \beta X_u$, we obtain
\begin{eqnarray}
\label{eq.pi_a}
\pi_a^\top &=&
\pi_a^\top Y_a^{-1} \Gamma_{a} X_a
+ \pi_u^\top \frac{X_u^{-1}}{\beta} \Gamma_{ua} X_a\\
\label{eq.pi_u}
\pi_u^\top &=&
\pi_a^\top Y_a^{-1} 
\Gamma_{au}
\left(I_{M_u} - \frac{\Gamma_u}{\beta}\right)^{-1}
X_u  \end{eqnarray}
In order to guarantee that the inverse appearing in \eqref{eq.pi_u} is positive and well-defined, it is necessary that $\beta > \beta_{min} =\lambda_F(\Gamma_u)$, in view of the properties of $M$-matrices \cite{berman94}. 
Substituting \eqref{eq.pi_u} into \eqref{eq.pi_a}, 
we finally get 
\begin{equation}
\label{eq.redcong}
\pi_a^\top  = \pi_a^\top Y_a^{-1}\Psi(\beta)X_a
\end{equation}
where
\begin{equation}
\label{eq.Psi}
\Psi(\beta)=\Gamma_a+\Gamma_{au}(\beta I_{M_u}-\Gamma_u)^{-1}\Gamma_{ua}, \quad Y_a={\rm diag}(\Psi(\beta)x_a)
\end{equation}
Note that \eqref{eq.redcong} can be interpreted as a kind of reduced-order version of the
congruence  equation \eqref{eq.congmat} with the real-valued positive matrix $\Psi(\beta)$ replacing the binary adjacency matrix $\Gamma$.
Notably, for a given $\beta$, the score vector $x_u$ can be directly obtained from $x_a$. Indeed,
$$
y_u=\Gamma_{ua} x_a + \Gamma_u x_u
$$
so that, recalling \eqref{eq.yi_u}, we have that
\begin{equation}
\label{eq.x_u}
x_u =(\beta I_{M_u} - \Gamma_u)^{-1}\Gamma_{ua} x_a
\end{equation}
In conclusion, the entropy maximization problem amounts to finding $\beta> \beta_{min}$ and a positive diagonal matrix $X_a$ that satisfy the reduced-order congruence equation \eqref{eq.redcong} and such that, by plugging \eqref{eq.x_u} into  \eqref{eq.pi_u}, the resulting vector $\begin{bmatrix}\pi_a^\top & \pi_u^\top\end{bmatrix}$ has unit sum. 
In this respect, note that, using \eqref{eq.pi_u}, \eqref{eq.Psi} and \eqref{eq.x_u}, it turns out that
\begin{equation}
\label{eq.gbeta}
g(\beta)=\begin{bmatrix}\pi_a^\top & \pi_u^\top\end{bmatrix} {\bf 1}_M = \pi_a^\top {\bf 1}_{M_a} + \beta \pi_a^\top \left(\diag(\Psi(\beta)x_a)\right)^{-1} \Gamma_{au} \left(\beta I_{M_u}-\Gamma_u\right)^{-2}\Gamma_{ua}x_a
\end{equation}
Thus, the optimum is reached when $g(\beta)=1$.
To solve this optimization problem, we propose the following iterative algorithm.
\medskip
\begin{algo}
\label{algo.2}
\,
\begin{itemize}
    \item {\bf Step 0.}
    $\Gamma, R$ and $\pi_a$ are given. 
    \item {\bf Step 1.}
    Let $\varepsilon_1, \varepsilon_2$ be two sufficiently small real positive scalars, let $\mu>1$ be a constant multiplier and set the maximum number of iterations $k_{max}$. Set $k=1$ and $\beta^{[0]}=\beta_{min} + \varepsilon_1$.
    \item {\bf Step 2.} If $k>k_{max}$, terminate the algorithm upon reaching the maximum number of iterations. Otherwise, run Algorithm \ref{algo.1} with inputs $\Gamma=\Psi(\beta^{[k]})$ and $\pi=\pi_a$. If it converges, call the solution $x_a$ and compute
    \begin{eqnarray*}
           x_u &=&\left(\beta^{[k]} I_{M_u} - \Gamma_u\right)^{-1}\Gamma_{ua} x_a \\
           y_a&=&\Psi(\beta^{[k]})x_a \\
    \pi_u^\top &=&
\pi_a^\top \left(\diag(y_a)\right)^{-1} 
\Gamma_{au}
\left(I_{M_u} - \frac{\Gamma_u}{\beta^{[k]}}\right)^{-1}
\diag(x_u)
    \end{eqnarray*}
else quit without a solution.
\item {\bf Step 3.} Evaluate
$$
g(\beta^{[k]})=\pi_a^\top {\bf 1}_{M_a} + \beta^{[k]} \pi_a^\top \left(\diag(y_a)\right)^{-1} \Gamma_{au} \left(\beta^{[k]} I_{M_u}-\Gamma_u\right)^{-2}\Gamma_{ua}x_a
$$
If $g(\beta^{[k]}) > 1$, let $\beta^{[k+1]} = \mu \beta ^{[k]}$, $k \leftarrow k+1$ and go to Step 2 else define $\beta_1=\beta^{[k-1]}$, 
$\beta_2=\beta^{[k]}$ and go to Step 4.

\item {\bf Step 4.} Find $\beta^*\in (\beta_1,\beta_2)$ such that $|\begin{bmatrix}
\pi_a^\top & \pi_u^\top
\end{bmatrix}
{\bf 1}_M - 1|<\varepsilon_2$ by a bisection procedure, using Algorithm 1. Letting $x_a$ be the final value returned by the algorithm in correspondence of $\beta^*$, compute
\begin{eqnarray*}
x_u &=&\left(\beta^* I_{M_u} - \Gamma_u\right)^{-1}\Gamma_{ua} x_a \\
y_a&=&\Psi(\beta^*)x_a \\
\pi_u^\top &=&
\pi_a^\top \left(\diag(y_a)\right)^{-1} 
\Gamma_{au}
\left(I_{M_u} - \frac{\Gamma_u}{\beta^*}\right)^{-1}
\diag(x_u) 
\end{eqnarray*}
and let  $\pi=R^\top \begin{bmatrix}
\pi_a^\top & \pi_u^\top
\end{bmatrix}^\top$,
$x=R^\top \begin{bmatrix}
x_a^\top & x_u^\top
\end{bmatrix}^\top$.
Finally, the maxentropic transition probability matrix $P$ and the associated entropy $H$ are obtained from \eqref{eq.paramat} and \eqref{eq.entr}, respectively.
\end{itemize}
\end{algo}
\medskip
The rationale behind Algorithm \ref{algo.2} is to adapt the value of the parameter $\beta$ until the sum $g(\beta)$ of all stationary probabilities in $\pi$ is equal to $1$. For $\beta$ tending to $\beta_{min}$ from above, the function $g(\beta)$ approaches infinity (thanks to irreducibility), while for $\beta$ tending to $+\infty$, it converges to a constant.
If this constant is less than $1$, there must exist a positive value of $\beta$ such that $g(\beta)=1$. In Steps 2 and 3 $\beta$ is progressively increased until a value of $\beta$ is found such that $g(\beta)<1$. At each step, the partial score vector $x_a$ of the assigned nodes is updated through Algorithm \ref{algo.1}. Then, in Step 4, a bisection search within the interval defined by the last two values of $\beta$ is performed in order to estimate the optimal value $\beta^*$, along with the associated optimal score vector $x$ and the complete stationary distribution $\pi$.

Algorithm \ref{algo.2} may fail to find the optimal solution compatible with the assigned partial distribution $\pi_a$ for one of the following two reasons.
First, it may happen that $g(\beta)>1$ for all $\beta>\beta_{min}$. In this case, the algorithm gets trapped in Steps 2 and 3 until the maximum number of iterations is reached, and no solution can be found. In our experiments, this typically occurs when $\pi_a$ is not assigned consistently with the given graph topology (e.g., a probability greater than or equal to $0.5$ is assigned to a node without a self-loop, see Theorem \ref{cor.onehalf}).
Second, Algorithm 1 invoked in Step 2 may not converge to a strictly positive partial score vector $x_a$, for some $\beta$. We conjecture that this happens for any value of $\beta$, again due to an inconsistent assignment of $\pi_a$ (for instance, leading to the violation of the constraint \eqref{eq.pi_consist}).

\begin{example}
\label{ex.partial}
Consider a 4-state Markov chain with adjacency matrix
$$
\Gamma = \begin{bmatrix}1 & 1 & 0 & 0 \\1 & 1 & 1 & 0 \\0 & 1 & 0 & 1 \\0 & 0 & 1 & 1\end{bmatrix}
$$
If the stationary distribution is completely free, the maxentropic solution is 
$\pi^*=\begin{bmatrix}0.2550 & 0.4686 & 0.1790 & 0.0974 \end{bmatrix}^\top$,
with $H^*=0.8568$ and
$$
P^*=\begin{bmatrix}
    0.4245 & 0.5755 & 0 & 0 \\
    0.3131 & 0.4245 & 0.2624 & 0 \\
    0 & 0.6869 & 0 & 03131 \\
    0 & 0 & 0.5755 & 0.4245
\end{bmatrix}
$$
Now assume that $\pi_1=0.1$ and $\pi_3=0.2$ are assigned. 
Running Algorithm 2 with $\epsilon_1=10^{-3}$, $\epsilon_2=10^{-6}$, $\mu=1.1$ and $k_{max}=50$, we obtained
the stationary distribution $\pi=\begin{bmatrix}
    0.1000 & 0.5102 & 0.2000 & 0.1898 
\end{bmatrix}^\top$ and the optimal transition probability matrix
$$
P=\begin{bmatrix}
    0.2058 & 0.7942 & 0 & 0 \\
    0.1557 & 0.6008 & 0.2435 & 0 \\
    0 & 0.6211 & 0 & 0.3789 \\
    0 & 0 & 0.3992 & 0.6008
\end{bmatrix}
$$
The associated entropy is $H=0.7906$.
\end{example}

\begin{remark}
Note that
$$
\lim_{\beta \to \infty} g(\beta)
=
2 {\bf 1}^\top \pi_a
-
\lim_{\beta \to \infty}
\pi_a^\top \diag(x_a)^{-1}
\left(
\Gamma_a
+
\frac{\Gamma_{au}\Gamma_{ua}}{\beta}
\right)^{-1}
\Gamma_a
x_a
$$
In particular, when $M_a=1$, namely when only one stationary probability is assigned, the reduced-order congruence equation \eqref{eq.redcong} is immediately satisfied letting $x_a=1$, for any $\beta$. Therefore, in Step 2  the execution of Algorithm \ref{algo.1} can be skipped. One is left with the problem of finding $\beta>\beta_{min}$ such that $g(\beta)=1$. In this case, 
$$
\lim_{\beta \to \infty} g(\beta)=\Biggl\{\begin{array}{cc}2 \pi_a &,\quad {\rm if}\quad \Gamma_a=0 \\\pi_a &,\quad {\rm if} \quad\Gamma_a=1 \end{array}
$$
We conclude that equation $g(\beta)=1$ always admits a solution when $\Gamma_a=1$, whereas if $\Gamma_a=0$ a necessary and sufficient condition for the existence of a solution is $\pi_a < 0.5$, in accordance with the necessary condition of Theorem \ref{cor.onehalf}.
\end{remark}

\section{Particular graph topologies}
In this section, we analyze three special classes of Markov chains which admit a simple characterization of the set $\mathcal D(\Gamma)$ of consistent steady- state probability distributions and a direct computation of the score vector associated with the maxentropic chain. 
\subsection{Complete graph with no self-loops}
Here we deal with the computation of the score vector associated with a complete graph without self-loops, i.e. $\Gamma_{ii}=0$, $\forall i$ and $\Gamma_{ij}=1$, $i\ne j$, in compact form $\Gamma={\bf 1}_M{\bf 1}_M^\top - I_M$. It  is clear that this graph is irreducible and aperiodic. Being symmetric, the normalized score of the maxentropic chain associated with any given consistent stationary probability vector is unique in the symplex, see Theorem \ref{th.symmetric} in the Appendix. If the steady-state probability distribution is free, the maxentropic chain is characterized by a uniform distribution $\pi = \frac{1}{M}\, {\bf 1}_M$. In the following statement, we characterize the set of consistent probability vectors. 
*** facile con Hall - precisare l'uso costruttivo della proof ***

\begin{theorem}
\label{th.complete}
Consider all Markov chains with adjacency matrix $\Gamma={\bf 1}_M{\bf 1}_M^\top - I_M$. Then, the set of consistent steady-state probabilities is $\mathcal{D}(\Gamma)=\left \{ \pi\in \Delta: \pi_i< 0.5, \forall i \right \}$.
\end{theorem}

\begin{proof}
In view of Theorem \ref{cor.onehalf}, if $\pi \in \mathcal{D}(\Gamma)$ then $\pi_i < 0.5, \forall i$. Therefore, we are left to prove that the inequality is also sufficient to ensure consistency. 
For this purpose, we will show that, under assumption $\pi_i< 0.5, \forall i$, it is possible to compute a score vector $x \in \Delta$, which  is associated 
through the parametrization \eqref{eq.x_param}
with a consistent steady-state probability distribution $\pi$.

First, observe that the graph is undirected because $\Gamma$ is symmetric.
Then, according to Proposition \ref{pro.symmetric}, any consistent $\pi$ must satisfy \eqref{eq.phi_sym} for some score vector $x \in \Delta$. Since $y = \Gamma x= {\bf 1}_M-x$, Eq. \eqref{eq.phi_sym} becomes 
$$
x_i(1-x_i)=\pi_i \sum_{k=1}^M x_k(1-x_k)
\quad , \quad \forall i
$$  
Letting $\alpha=\sum_{k=1}^M x_k(1-x_k)$, the single scores are then given by 
\begin{equation}
\label{eq.score_complete}    
x_i=1/2 \pm \sqrt{1/4-\alpha\pi_i} \quad , \quad \forall i
\end{equation}
In order for all scores to be well defined, it is necessary that $\alpha\le \alpha_{max}=\frac{1}{4\pi_{max}}$, where $\pi_{max}=\max_i \{\pi_i\}$. Moreover, at most only one of the signs in front of the square roots in \eqref{eq.score_complete} is plus, otherwise the constraint $\sum_i x_i=1$ is not met. Also, notice that all nodes in the graph are directly connected and topologically indistinguishable. Hence, the node with the highest score must have a steady-state probability $\pi_{max}$. If there are multiple nodes with the same steady-state probability $\pi_{max}$, they must have the same score, so that all signs in \eqref{eq.score_complete} are necessarily minus. Let $i_{max}$ be any index $i$ such that $\pi_i=\pi_{max}$ 
and consider the scalar functions
\begin{eqnarray*}
f(\alpha)&=&M/2+\sqrt{1/4-\alpha\pi_{max}}-\sum_{i\ne i_{max}} \sqrt{1/4-\alpha\pi_i}\\
g(\alpha)&=&M/2-\sum_{i} \sqrt{1/4-\alpha\pi_i}
\end{eqnarray*}
The function $f(\alpha)$ represents the sum of the scores $x_i$ when there is a single plus sign in front of the square roots in \eqref{eq.score_complete}, while the function $g(\alpha)$ represents the sum of the scores when all signs are minus.
In order to obtain a score vector $x \in \Delta$ associated with a consistent steady-state probability distribution, either $f(\alpha)$ or $g(\alpha)$ must be equal to $1$.
Notice that $f(0)=1$, $g(0)=0$, and $f(\alpha_{max})=g(\alpha_{max})=\frac{1}{2}\left(M-\sum_{i\ne i_{max}}\sqrt{1-\pi_i/\pi_{max}}\right)$. 
Letting $f'(\alpha)$ and $g'(\alpha)$ be the derivatives of the two functions, we have 
$g'(\alpha)>0$, for $\alpha \in[0,\alpha_{\mathrm{max}}]$, while $f'(\alpha)$ is positive (thanks to the initial assumption $\pi_i<0.5, \forall i$, or equivalently $\pi_{max}<0.5$) for $\alpha\in [0, \alpha^*)$ and negative for $\alpha\in (\alpha^*,\alpha_{max})$, being zero in $\alpha^*$, where there is a maximum. Thus, we conclude that, 
if $g(\alpha_{max})\ge 1$, 
there exists a value $\bar\alpha \in (0,\alpha_{max}]$ such that
$g(\bar\alpha)=1$ and the associated scores are $x_i=1/2-\sqrt{1/4-\bar\alpha \pi_i}, \forall i$. On the contrary, if $g(\alpha_{max})<1$, there exists a value of $\bar\alpha \in (0,\alpha_{max})$ such that $f(\bar\alpha)=1$ and the associated scores are 
$x_{i_{max}}=1/2+\sqrt{1/4-\bar\alpha \pi_{max}}$ and $x_{i}=1/2-\sqrt{1/4-\bar\alpha \pi_i}, i\ne i_{max}$.
In conclusion, any steady-state probability distribution $\pi$ with $\pi_i<0.5, \forall i$, can be determined from a suitable score vector $x \in \Delta$. As such, it is consistent with the given graph topology.   
\end{proof}

Notice that the value of $\bar\alpha$ defined in the proof can be numerically computed by a line search in a finite interval.
The procedure for computing the normalized score $x$ illustrated in the proof provides an alternative way to Algorithm \ref{algo.1} to obtain the maxentropic chain in this special case of a complete graph with no self-loops. 

\begin{example}
\medskip\noindent

    (i) Let $M=3$, $\Gamma={\bf 1}_3{\bf 1}_3^\top - I_3$ and $\pi=\frac{1}{11}[4 \quad 3 \quad 4]^\top$. The curves $f(\alpha)$ and $g(\alpha)$ for $\alpha \in (0,\alpha_{max})$, with $\alpha_{max}=0.6875$, are reported in Figure \ref{fig.complete_1}. It is apparent that $g(\alpha_{max})>1$, so that the value of $\bar\alpha$ is obtained from the intersection of the curve $g(\alpha)$ with the line with ordinate $1$. From a line search, it results $\bar\alpha=0.6509$. The associated scores are
    $x_i=1/2-\sqrt{1/4-\bar\alpha \pi_i},  i=1,2,3$, i.e.
    $$
    x = \begin{bmatrix}
    0.3846 & 0.2308 & 0.3846
    \end{bmatrix}^\top
    $$
    the maxentropic transition probability matrix is
    $$
    P = \begin{bmatrix}
         0  &  0.3750  &  0.6250 \\
    0.5000  &       0  &  0.5000 \\
    0.6250  &  0.3750  &       0
    \end{bmatrix}
    $$
    and the maximum entropy is $H^*=0.6702$.
\begin{figure}
        \centering
        \includegraphics[width=8cm]{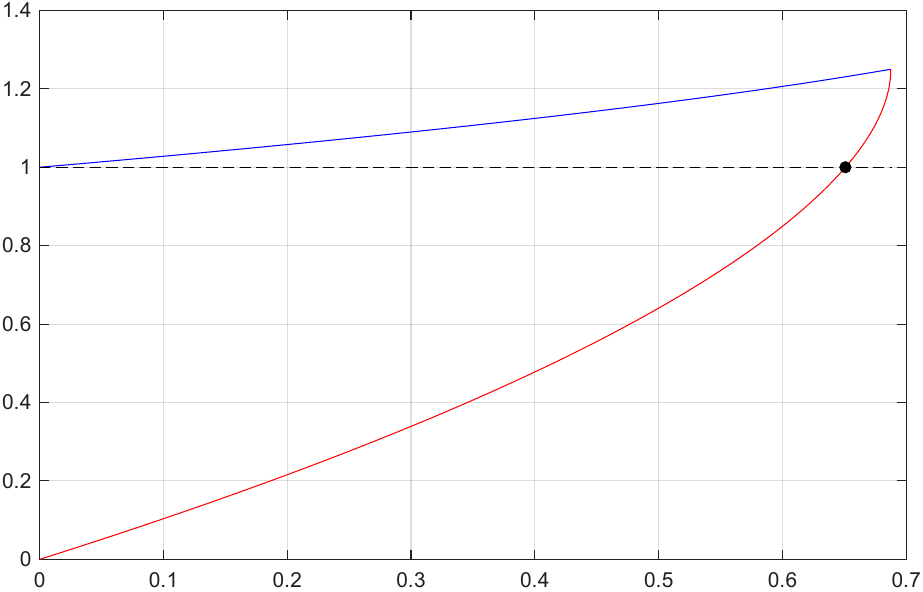}
        \caption{Case (i)}
        \label{fig.complete_1}
\end{figure}
    
    \medskip\noindent

    (ii) Let $M=3$, $\Gamma={\bf 1}_3{\bf 1}_3^\top - I_3$ and $\pi=\frac{1}{11}[5 \quad 2 \quad 4]^\top$. The curves $f(\alpha)$ and $g(\alpha)$ for $\alpha \in (0,\alpha_{max})$, with $\alpha_{max}=0.5500$, are reported in Figure \ref{fig.complete_2}. It is apparent that $g(\alpha_{max})<1$, so that the value of $\bar\alpha$ is obtained from the intersection of the curve $f(\alpha)$ with the line with ordinate $1$. From a line search, it results $\bar\alpha=0.4808$. Since $i_{max}=1$, the associated scores are
    $x_1=1/2+\sqrt{1/4-\bar\alpha \pi_1}$ and $x_i=1/2-\sqrt{1/4-\bar\alpha \pi_i}, i=2,3$, i.e.
    $$
    x = \begin{bmatrix}
     0.6774  &  0.0968  &  0.2258
    \end{bmatrix}^\top
    $$
    the maxentropic transition probability matrix is
    $$
    P = \begin{bmatrix}
         0  &  0.3000  &  0.7000 \\
    0.7500  &       0  &  0.2500 \\
    0.8750  &  0.1250  &       0
    \end{bmatrix}
    $$
    and the maximum entropy is $H^*=0.5169$.
\begin{figure}
        \centering
        \includegraphics[width=8cm]{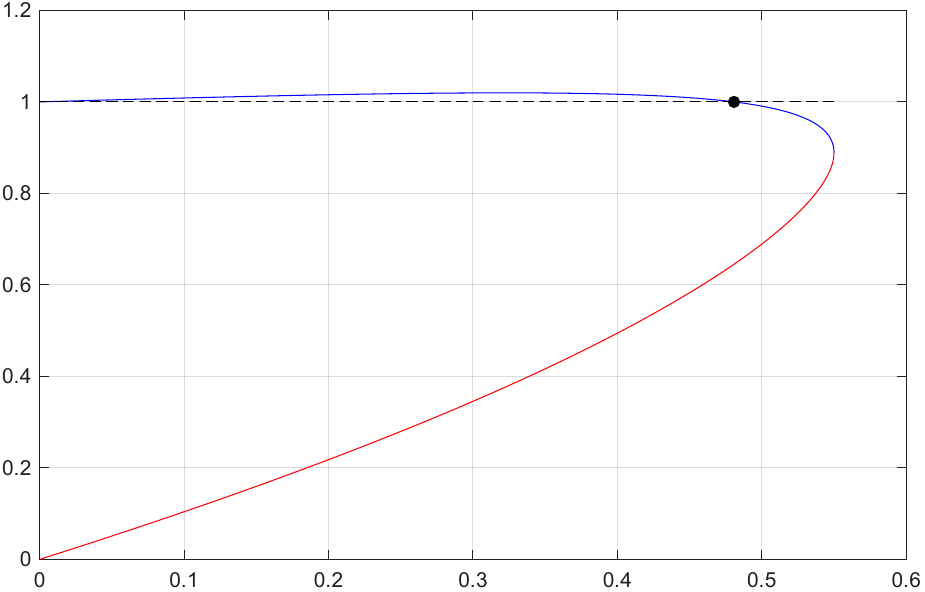}
        \caption{Case (ii)}
        \label{fig.complete_2}
\end{figure} 
\end{example}

\subsection{String graph with a single self-loop}
Consider an undirected graph with string topology where only one node (say the $k$-th node) has a self-loop. Then the adjacency matrix has this form:

\begin{equation}
\label{eq.string}
\Gamma =
\begin{bNiceArray}{ccccccccc}[first-row, last-col]
    & & & & \substack{\textstyle k \\ \textstyle \downarrow}\\
    0 & 1 & 0 & \cdots & \cdots & \cdots & 0 & 0 & 0\\
    1 & 0 & 1 & \cdots & \cdots & \cdots & 0 & 0 & 0\\
    0 & 1 & 0 & \cdots & \cdots & \cdots & 0 & 0 & 0\\
    \vdots & \vdots & \vdots & \ddots & \vdots & \vdots
        & \vdots & \vdots & \vdots\\
    \vdots & \vdots & \vdots & 1 & 1 & 1
        & \vdots & \vdots & \vdots & \leftarrow k\\
    \vdots & \vdots & \vdots & \vdots & \vdots & \ddots
        & \vdots & \vdots & \vdots\\
    0 & 0 & 0 & \cdots & \cdots & \cdots & 0 & 1 & 0\\
    0 & 0 & 0 & \cdots & \cdots & \cdots & 1 & 0 & 1\\
    0 & 0 & 0 & \cdots & \cdots & \cdots & 0 & 1 & 0
\end{bNiceArray}
\end{equation}    
This chain is irreducible and aperiodic (thanks to the existence of a self-loop). Since $\Gamma$ is symmetric, the normalized score vector of the maxentropic chain associated with any given consistent stationary probability vector is unique in the symplex. The next theorem provides a characterization of the set of consistent probability vectors along with a procedure to compute the scores associated with the maxentropic chain.
The following index subsets notation will be used:
\begin{eqnarray*}
    \mathcal{J}_{Oi}^- &=& \{j \in \mathcal M, j<i, (j-i) \,\, {\mathrm {odd}}\} \\
    \mathcal{J}_{Ei}^- &=& \{j \in \mathcal M, j<i, (j-i) \,\, {\mathrm {even}}\}
    \\
    \mathcal{J}_{Oi}^+ &=& \{j \in \mathcal M, j>i, (j-i) \,\, {\mathrm {odd}}\} \\
    \mathcal{J}_{Ei}^+ &=& \{j \in \mathcal M, j>i, (j-i) \,\, {\mathrm {even}}\}
    \\
    \mathcal{J}_{Oi} &=& \mathcal{J}_{Oi}^- \cup \mathcal{J}_{Oi}^+
    \\
    \mathcal{J}_{Ei} &=& \mathcal{J}_{Ei}^- \cup \mathcal{J}_{Ei}^+
\end{eqnarray*}

\begin{theorem}
\label{th.string}
Consider all Markov chains with adjacency matrix $\Gamma$ with the structure \eqref{eq.string}. Then, the set $\mathcal D(\Gamma)$ of consistent steady-state probabilities is defined by the following inequalities:
\begin{eqnarray}
\pi_i &>& \sum_{j \in \mathcal {J}_{Oi}^-}\pi_j
-\sum_{j \in \mathcal {J}_{Ei}^-}\pi_j
\quad , \quad i \in [2,k-1]
\label{eq.constr_string1}\\
\pi_i &>& \sum_{j \in \mathcal {J}_{Oi}^+}\pi_j
-\sum_{j \in \mathcal {J}_{Ei}^+}\pi_j
\quad , \quad i \in [k+1,M-1]
\label{eq.constr_string2}\\
\pi_i &>& \sum_{j \in \mathcal {J}_{Oi}}\pi_j
-\sum_{j \in \mathcal {J}_{Ei}}\pi_j
\quad , \quad i=k
\label{eq.constr_string3}
\end{eqnarray}
Moreover, the score vector of the maxentropic chain is recursively computed from the equations
\begin{eqnarray}
x_i &=& \sqrt{u_i} \quad,\quad i=k 
\label{eq.score_string1}\\
x_i &=& \frac{u_i}{x_{i+i}} \quad,\quad i=k-1,k-2,\ldots,1
\label{eq.score_string2}\\ 
 x_i &=& \frac{u_i}{x_{i-i}} \quad,\quad i=k+1,k+2, \ldots,M
 \label{eq.score_string3}
\end{eqnarray}
where 
\begin{eqnarray}
u_i &=& \pi_i - \sum_{j \in \mathcal {J}_{Oi}^-}\pi_j
+\sum_{j \in \mathcal {J}_{Ei}^-}\pi_j
\quad , \quad i \in [1,k-1]
\label{eq.cong_string1}\\
u_i &=& \pi_i - \sum_{j \in \mathcal {J}_{Oi}^+}\pi_j
+\sum_{j \in \mathcal {J}_{Ei}^+}\pi_j
\quad , \quad i \in [k+1,M]
\label{eq.cong_string2}\\
u_i &=& \pi_i - \sum_{j \in \mathcal {J}_{Oi}}\pi_j
+\sum_{j \in \mathcal {J}_{Ei}}\pi_j
\quad , \quad i=k
\label{eq.cong_string3}
\end{eqnarray}
\end{theorem}

\begin{proof}
Note that $\Gamma$ is symmetric.
Then, according to Proposition \ref{pro.symmetric}, any consistent $\pi$ must satisfy the congruence equation \eqref{eq.phi_sym} for some score vector $x$, namely
\begin{eqnarray*}
x_1 x_2 &=& \alpha \pi_1\\
x_2 (x_1+x_2) &=& \alpha \pi_2\\
&\cdots& \\
x_k(x_{k-1}+x_k+x_{k+1}) &=& \alpha \pi_k\\
&\cdots& \\
x_{M-1} (x_{M-2}+x_M) &=& \alpha \pi_{M-1}\\
x_{M-1} x_M &=& \alpha \pi_M
\end{eqnarray*}
with $\alpha = x^\top \Gamma x$. Without loss of generality, we can put $\alpha=1$, using a different normalization of $x$ rather than assuming $x \in \Delta$.
In terms of the positive auxiliary variables 
\begin{eqnarray}
u_i &=& x_i x_{i+1} \quad,\quad i \ne k
\label{eq.u_string1}\\
u_k &=& x_k^2
\label{eq.u_string2}
\end{eqnarray}
the set of congruence equations can be written in linear form as 
\begin{eqnarray*}
u_1 &=& \pi_1\\
u_1+u_2 &=& \pi_2\\
u_2+u_3 &=& \pi_3\\
&\cdots& \\
u_{k-1}+u_k+u_{k+1} &=& \pi_k\\
&\cdots& \\
u_{M-3}+u_{M-2} &=& \pi_{M-2}\\
u_{M-2}+u_{M-1} &=& \pi_{M-1}\\
u_M &=& \pi_M
\end{eqnarray*}
It is simple to show that the solution of such a set of linear equations is given by \eqref{eq.cong_string1}-\eqref{eq.cong_string3}. 
The positivity of all variables $u_i$ (and, in turn, the consistency of a given $\pi$) is ensured by the constraints \eqref{eq.constr_string1}-\eqref{eq.constr_string3}.
Finally, the expression of the 
scores \eqref{eq.score_string1}-\eqref{eq.score_string3} is obtained recursively by inverting equations \eqref{eq.u_string1}-\eqref{eq.u_string2}.
\end{proof}
\begin{example}
Consider a Markov chain with an undirected string graph with $M=5$ nodes and a self-loop only in node $k=2$.
If the steady-state probability distribution is free, the maxentropic chain has 
$$
\pi=[0.1199 \quad 0.5423 \quad 0.2338 \quad 0.0852 \quad   0.0188]^\top
$$
and the optimal vector score is 
$$
x=[0.1735 \quad 0.3691 \quad 0.2423 \quad 0.1463 \quad 0.0688]^\top
$$
The maximum value of the entropy is $H^*=0.7546$. 
Now, assume that the steady-state probability distribution 
$$
\pi=[0.1 \quad 0.4 \quad 0.2 \quad 0.2 \quad   0.1]^\top
$$
is assigned. Using the procedure described above, we find that such a distribution is actually consistent, the unknown auxiliary vector is $u=[0.1 \quad 0.2 \quad 0.1 \quad 0.1 \quad 0.1]^\top$, and the (unnormalized) score vector is 
$$
x=[0.2236 \quad  0.4472 \quad 0.2236 \quad 0.447 \quad 0.2236]^\top
$$
The corresponding normalized score vector $\bar x \in\Delta$ is then $\bar x=[0.1429 \quad 0.2857 \quad 0.1429 \quad 0.2857 \quad 0.1429]^\top$.
The transition probability matrix
is given by 
$$
P = \begin{bmatrix}
         0  &  1.0  &  0 & 0 & 0
         \\ 0.25 & 
    0.5  &  0.25 & 0 & 0 \\
    0 & 0.5 & 0 & 0.5 & 0 \\
    0 & 0 & 0.5 & 0 & 0.5 \\
0 & 0 & 0 & 1 & 0
    \end{bmatrix}
    $$
and the associated entropy is $H=0.6931$.
\end{example}

\subsection{Odd ring graph without self-loops}
Consider an undirected graph with ring topology with no self-loops and an odd number of nodes. The corresponding Markov chain is irreducible and aperiodic, thanks to the assumption of an odd number of nodes. The associated adjacency matrix has the following structure:
\begin{equation}
\label{eq.oddring}
\Gamma =
\begin{bmatrix}
         0  &  1  &  0 &  \cdots & 0 & 0 & 1\\
    1  & 0 & 1 & \cdots & 0 & 0 & 0\\
    0 & 1 & 0 & \cdots &0 & 0 & 0 \\
    \vdots  & \vdots & \vdots  & \ddots & \vdots & \vdots & \vdots \\
    0 & 0 & 0 & \cdots & 0 & 1 & 0\\
    0 & 0 & 0 & \cdots& 1 & 0 &1\\
    1 & 0 & 0 & \cdots &0 & 1 & 0
\end{bmatrix}
\end{equation}
Since $\Gamma$ is  symmetric, the normalized score vector of the maxentropic chain associated with any given consistent stationary probability vector is unique in the symplex. The following theorem provides a characterization of the set of consistent probability vectors along with a procedure to compute the scores associated with the maxentropic chain. In the statement, the symbol $\lfloor q \rfloor$ denotes the integer part of the positive real number $q$.
\begin{theorem}
\label{th.oddring}
Consider all Markov chains with an odd $M$ and an adjacency matrix $\Gamma$ with the structure \eqref{eq.oddring}. Then, the set $\mathcal D(\Gamma)$ of consistent steady-state probabilities is defined by the following inequalities, for $i=1,2,\ldots,M$:
\begin{equation}
\label{eq.constr_oddring}
0.5<\pi_i + \sum_{j=1}^{\left\lfloor\frac{M}{2}\right\rfloor}
\pi_{k(i,j)}\quad,\quad
k(i,j)=i+2j-1-M \left\lfloor\frac{i+2j-2}{M}\right\rfloor
\end{equation}
Moreover, the score vector of the maxentropic chain can be obtained from the following equations:
\begin{eqnarray}
x_1 &=& \sqrt{\frac{\prod_{j=0}^{\lfloor \frac{M}{2}\rfloor} u_{2j+1}}
{\prod_{j=1}^{\lfloor \frac{M}{2}\rfloor} u_{2j}}}
\label{eq.score_oddring1}\\
 x_i &=& \frac{u_{i-1}}{x_{i-i}} \quad,\quad i=2,3,\ldots,M
 \label{eq.score_oddring2}
\end{eqnarray}
where 
\begin{equation}
u_i = 0.5 \left(
\pi_i + \sum_{j=1}^{\left\lfloor\frac{M}{2}\right\rfloor}
\left(
\pi_{k(i,j)} - \pi_{h(i,j)}
\right)
\right)
\quad, \quad i=1,2,\ldots,M
\label{eq.cong_oddring}
\end{equation}
and the indices $k(i,j)$, $h(i,j)$ are defined as
\begin{equation}
\label{eq.ind_oddring}
k(i,j)=i+2j-1-M \left\lfloor\frac{i+2j-2}{M}\right\rfloor
\quad,\quad 
h(i,j)=i+2j-M \left\lfloor\frac{i+2j-1}{M}\right\rfloor
\end{equation}
\end{theorem}
\begin{proof}
The proof is similar to the proof of Theorem \ref{th.string}. After defining the auxiliary variables 
\begin{eqnarray}
u_i &=& x_i x_{i+1} \quad,\quad i=1,2,\ldots, (M-1)
\label{eq.u_oddring1}\\
u_M &=& x_M x_1
\label{eq.u_oddring2}
\end{eqnarray}
it is easy to observe that they solve the congruence equations \eqref{eq.cong_oddring}, \eqref{eq.ind_oddring}.
The consistency conditions \eqref{eq.constr_oddring} are obtained by imposing the positivity of these auxiliary variables. From the vector $u$, the score vector is computed by inverting \eqref{eq.u_oddring1}, \eqref{eq.u_oddring2} through the iterative eqations \eqref{eq.score_oddring1}, \eqref{eq.score_oddring2}. 
\end{proof}

\begin{example}
Consider a Markov chain with an undirected ring graph with $M=5$ nodes and no self-loops.
If the steady-state probability distribution is free, the maxentropic chain obviously has a uniform distribution $\pi=0.2\,{\bf 1}_5$, the scores are all equal, and the maximum value of the entropy is 
$H^*=0.6931$. 
The steady-state probability distribution 
$$
\pi=[0.20 \quad 0.25 \quad 0.15 \quad 0.20 \quad  0.20]^\top
$$
satisfies conditions \eqref{eq.constr_oddring}, so it 
is consistent. By solving equations \eqref{eq.u_oddring1}, \eqref{eq.u_oddring2},
one obtains the auxiliary vector 
$$
u=[0.15 \quad 0.10 \quad 0.05 \quad 0.15 \quad 0.05]^\top
$$
The (unnormalized) score vector is 
$$
x=[0.1581 \quad  0.9487 \quad 0.1054 \quad 0.4743 \quad 0.3162]^\top
$$
The corresponding normalized score vector $\bar x \in\Delta$ is 
$$
\bar x=[0.0789 \quad 0.4737 \quad 0.0526 \quad 0.2368 \quad 0.1579]^\top
$$
The transition probability matrix
is given by 
$$
P = \begin{bmatrix}
         0  &  0.75  &  0 & 0 & 0.25
         \\ 0.60 & 
    0  &  0.40 & 0 & 0 \\
    0 & 0.67 & 0 & 0.33 & 0 \\
    0 & 0 & 0.25 & 0 & 0.75 \\
0.25 & 0 & 0 & 0.75 & 0
    \end{bmatrix}
    $$
and the associated entropy is $H=0.6011$.
\end{example}

\section{Concluding remarks}

\section*{Appendix}

\subsection*{Nonuniqueness of the normalized score vector}

The normalized score vector $x$ is in general not unique for an irreducible and aperiodic maxentropic Markov chain.
The following theorem provides a necessary and sufficient condition for nonuniqueness. 

\begin{theorem}
\label{th.Delta}
A necessary and sufficient condition for the nonuniqueness of the normalized score vector of a maxentropic chain is the existence of two positive diagonal matrices $\Delta_1$, $\Delta_2$, both not proportional to the identity matrix, such that 
\begin{equation}
    \label{eq.Delta}
\Gamma=\Delta_1\Gamma\Delta_2
\end{equation}
\end{theorem}
\begin{proof} Let $x$ be a score that satisfies \eqref{eq.congmat} for a given $\pi$, where $y=\Gamma x$, and consider the associated transition probability matrix $P$ in \eqref{eq.paramat}.  Assume that \eqref{eq.Delta} is satisfied for some positive diagonal matrices $\Delta_1, \Delta_2$, and take $z=\Delta_2 x$ (which is not proportional to $x$). Finally, take $w=\Gamma z$ and compute $\bar P=W^{-1}\Gamma Z$, where $W={\rm diag}(w)$ and $Z={\rm diag}(z)$. It turns out that
$$
\bar P=W^{-1}\Gamma Z={\rm diag}(\Gamma \Delta_2 x)^{-1}\Gamma \Delta_2 X={\rm diag}(\Delta_1^{-1}\Gamma x)^{-1}\Gamma \Delta_2 X={\rm diag}(\Gamma x)^{-1}\Delta_1\Gamma \Delta_2 X=Y^{-1}\Gamma X=P
$$
Therefore, $z$ is a score vector, not proportional to $x$, yielding the same Markov chain.

Viceversa, assume that the normalized $x$ satisfying \eqref{eq.congmat} is not unique, and let $z=\Delta_2x$ be another score (not proportional to $x$) generating the same transition probability matrix $P$. Also, let $w=\Gamma z$. Then, reversing the previous arguments, we have $\Gamma=\Delta_1\Gamma \Delta_2$ with $\Delta_1=YW^{-1}$. Note that $\Gamma=\Delta_1\Gamma \Delta_2$ implies that $\Delta_1$ cannot be proportional to the identity matrix, so the proof is concluded. \end{proof}

\medskip\noindent

Note that the number of distinct values appearing in the diagonal of $\Delta_1$ represents the number of independent scaling factors that can be applied to subvectors of $x$ without changing the associated transition probability $P$. The maximum number of such scaling factors is a property of the adjacency matrix $\Gamma$ and will be called L-dof (Luce's degree of freedom). Next, before discussing how to find the L-dof of a given graph, we provide an equivalent condition for nonuniqueness of the normalized score vector. 

\begin{theorem}
\label{th.antidiag}
A necessary and sufficient condition for the nonuniqueness of the normalized score vector of a maxentropic chain with adjacency matrix $\Gamma$ is the existence of a permutation matrix $Q$ such that $Q \Gamma Q^\top$ is block antidiagonal. 
\end{theorem}
\begin{proof}
Assume that a permutation matrix $Q$ exists such that 
$$
Q\Gamma Q^\top:=\bar \Gamma=\left[ \begin{array}{ccccc} 
0 & 0 & \cdots & 0 & \Gamma_1\\
0 & 0 & \cdots &\Gamma_2 & 0\\
\vdots & \vdots & \cdot^{\cdot^\cdot} & \vdots & \vdots \\
0 & \Gamma_{k-1} & \cdots & 0 & 0\\
\Gamma_k & 0 & \cdots & 0 & 0
\end{array}\right]
$$
where $\Gamma_i$ are square matrices of $0$ and $1$ with dimensions $n_i$, $i=1,2,\cdots, k$. 
Then, it is a matter of simple computation to show that 
$$
\Delta_1=Q^\top {\rm blkdiag}\{\alpha_i I_{n_i}\}Q \quad, \quad \Delta_2=Q^\top {\rm blkdiag}\{\alpha_{k-i+1}^{-1} I_{n_{k-i+1}}\}Q
$$
satisfy \eqref{eq.Delta}.
Hence, if $x$ is a score vector, then $z=\Delta_2 x$ is an equivalent score vector for any set of distinct $\alpha_i\in R_+, i=1, \ldots, k$. 

Viceversa, if the normalized $x$ is not unique, then, recalling Theorem \ref{th.Delta}, consider all possible $\Delta_1, \Delta_2$ such that $\Gamma=\Delta_1\Gamma\Delta_2$ and notice that $\Delta_1$ and $\Delta_2$ are linked by the equations $\Delta_1(i,i)\Delta_2(j,j)=1$, for any $(i,j)$ such that $\Gamma_{ij}\ne 0$. Matrix $\Delta_1$ is composed by $k\le M$ distinct positive diagonal entries $\alpha_i$, each one repeated $n_i$ times, associated with a partition $U_i$, $i=1,2,\cdots, k$ of the rows of $\Gamma$. Letting $Q$ be the permutation matrix such that 
$Q\Delta_1Q^\top={\rm blkdiag}\{\alpha_i I_{n_i}\}$, it is clear that 
$$
Q \Delta_2 Q^\top ={\rm blkdiag}\{\alpha_{k-i+1}^{-1} I_{n_{k-i+1}}\}
$$
Hence, letting $\bar \Gamma=Q\Gamma Q^\top$, we have that
$$
\bar \Gamma ={\rm blkdiag}\{\alpha_i I_{n_i}\} \, \bar \Gamma \, {\rm blkdiag}\{\alpha_{k-i+1}^{-1} I_{n_{k-i+1}}\}.
$$
Recalling that all $\alpha_i$ are distinct, then $\bar \Gamma$ is necessarily block-antidiagonal.
\end{proof}

The degree of freedom L-dof$(\Gamma)$ corresponds to the maximum number of distinct $\alpha_i$ and is therefore equal to the maximum number of antidiagonal blocks. As observed in Example \ref{ex.nonunique}, the column indices of each antidiagonal block identify a disjoint state subsets $\mathcal A_i, i=1, \ldots, \rm{L-dof}(\Gamma)$, such that $\mathcal A_i=\Gamma_{out}(\Gamma_{in}(\mathcal A_i))$.

\begin{example}
Let
$$
\Gamma=
\left[\begin{array}{ccccc}
0  &   0  &   1  &   1  &   1 \\
0  &   0  &   1  &   0  &   1 \\
0  &   0  &   1  &   0  &   1 \\
0  &   1  &   0  &   0  &   0 \\
1  &   0  &   0  &   0  &   0
\end{array}\right]
$$
This graph is irreducible and aperiodic. However, the score associated with any feasible stationary distribution $\pi$ is not unique in the symplex. Indeed, $\Gamma$ s already in block antidiagonal form, with  
$$
\Gamma_1=
\left[\begin{array}{ccc}
1  &   1 & 1 \\
1  &   0 & 1\\
1 & 1 & 0 
\end{array}\right], \quad \Gamma_2=1, \quad \Gamma_3=1
$$
so that $\Gamma=\Delta_1\Gamma \Delta_2$, with 
$$
\Delta_1=
\left[\begin{array}{ccccc}
\alpha_1  &   0 & 0 & 0 & 0  \\
0 & \alpha_1  &   0 & 0 & 0\\
0 & 0 & \alpha_1 & 0 & 0\\
0 & 0 & 0 & \alpha_2 & 0\\
0 & 0 & 0 & 0 &\alpha_3
\end{array}\right], \quad \Delta_2=\left[\begin{array}{ccccc}
\alpha_3^{-1}  &   0 & 0 & 0 & 0  \\
0 & \alpha_2^{-1}  &   0 & 0 & 0\\
0 & 0 & \alpha_1^{-1} & 0 & 0\\
0 & 0 & 0 & \alpha_1^{-1} & 0\\
0 & 0 & 0 & 0 &\alpha_1^{-1}
\end{array}\right]
$$
Hence, if $x$ is a score associated with a given $\pi$, all other scores associated with the same $\pi$ and generating the same $P$ are given by $z=k\Delta_2 x$, for any positive $\alpha_1, \alpha_2, \alpha_3, k$ (normalized). In this example, it is seen that $\mathcal A_i=\Gamma_{out}(\Gamma_{in}(\mathcal A_i))$, for the three subsets $A_1=\{1\}, A_2=\{2\}, A_3=\{3,4,5\}$.
\end{example}

The next theorem shows that for a maxentropic chain with undirected graph (i.e with symmetric adjacency matrix) the normalized score vector is unique.
\begin{theorem}
\label{th.symmetric}
Assume that the adjacency matrix $\Gamma$ is irreducible, aperiodic and symmetric. Then the normalized score vector of the associated maxentropic Markov chain is unique.  
\end{theorem}
\begin{proof}
The proof is by contradiction. If the score vector is not unique, then, in view of Theorem \ref{th.antidiag}, there exists a permutation matrix $Q$ such that $\bar\Gamma=Q\Gamma Q^\top$ is block antidiagonal. Note that $\bar\Gamma$ is irreducible, aperiodic and symmetric.
Let $M$ be the size of $\Gamma$ and $k$ be the number of antidiagonal blocks $\bar\Gamma_i$ of $\bar\Gamma$. If $M$ is even, the block $\bar\Gamma_i$ must have the same dimension as the block $\bar\Gamma_{k-i+1}$. Then, the power $\bar\Gamma^n$ contains zero entries for any positive integer $n$. Hence, $\bar\Gamma$ is not aperiodic, a contradiction. 
If $M$ is odd, also $k$ is odd and the row indices of the middle block $\bar\Gamma_{(k+1)/2}$ coincide with its column indices, so that this block corresponds to an isolated set of states, not communicating with the other states. Hence, $\bar\Gamma$ violates the assumption of irreducibility. 
\end{proof}

\bibliographystyle{plain}
\bibliography{Markov26}

@book{Berman94,
  author       = {Berman, Abraham and Plemmons, Robert J.},
  title        = {Nonnegative matrices in the mathematical sciences},
  publisher    = {SIAM},
  volume       = {9},
  year         = {1994},
}

@article{BCDAutomatica19,
  author       = {Bolzern, Paolo and Colaneri, Patrizio and De Nicolao, Giuseppe},
  title        = {Opinion influence and evolution in social networks: A {M}arkovian agents model},
  journal      = {Automatica},
  publisher    = {Elsevier},
  volume       = {100},
  pages        = {219--230},
  year         = {2019},
}

@article{BCD_Automatica25,
  author       = {Bolzern, Paolo and Colaneri, Patrizio and De Nicolao, Giuseppe},
  title        = {Maxentropic continuous-time homogeneous {M}arkov chains},
  journal      = {Automatica},
  publisher    = {Elsevier},
  volume       = {175},
  pages        = {112214},
  year         = {2025},
}

@article{Bolker72,
  author       = {Bolker, Ethan D.},
  title        = {Transportation Polytopes},
  journal      = {Journal of Combinatorial Theory, Series B},
  volume       = {13},
  number       = {3},
  pages        = {251--262},
  year         = {1972},
  doi          = {10.1016/0095-8956(72)90060-3},
}

@article{Brualdi68,
  author       = {Brualdi, Richard A.},
  title        = {Convex sets of non-negative matrices},
  journal      = {Canadian Journal of Mathematics},
  publisher    = {Cambridge University Press},
  volume       = {20},
  pages        = {144--157},
  year         = {1968},
}

@article{Bullo19,
  author       = {George, Mishel and Jafarpour, Saber and Bullo, Francesco},
  title        = {Markov chains with maximum entropy for robotic surveillance},
  journal      = {IEEE Transactions on Automatic Control},
  publisher    = {IEEE},
  volume       = {64},
  number       = {4},
  pages        = {1566--1580},
  year         = {2019},
}

@article{Burda09,
  author       = {Burda, Zdzislaw and Duda, Jarek and Luck, Jean-Marc and Waclaw, Bartek},
  title        = {Localization of the maximal entropy random walk},
  journal      = {Physical Review Letters},
  publisher    = {APS},
  volume       = {102},
  number       = {16},
  pages        = {160602},
  year         = {2009},
}

@book{Cassandras08,
  author       = {Cassandras, Christos G. and Lafortune, St{\'e}phane},
  title        = {Introduction to discrete event systems},
  publisher    = {Springer},
  year         = {2008},
}

@article{Delvenne11,
  author       = {Delvenne, Jean-Charles and Libert, Anne-Sophie},
  title        = {Centrality Measures and Thermodynamic Formalism for Complex Networks},
  journal      = {Physical Review E},
  volume       = {83},
  number       = {4},
  pages        = {046117},
  year         = {2011},
  doi          = {10.1103/PhysRevE.83.046117},
}

@phdthesis{Duda12,
  author       = {Duda, Jaroslaw},
  title        = {Extended maximal entropy random walk},
  school       = {Jagiellonian University},
  year         = {2012},
}

@article{Franklin89,
  author       = {Franklin, Joel and Lorenz, Jens},
  title        = {On the Scaling of Multidimensional Matrices},
  journal      = {Linear Algebra and its Applications},
  volume       = {114--115},
  pages        = {717--735},
  year         = {1989},
  doi          = {10.1016/0024-3795(89)90490-4},
}

@article{Gale57,
  author       = {Gale, David},
  title        = {A Theorem on Flows in Networks},
  journal      = {Pacific Journal of Mathematics},
  volume       = {7},
  number       = {2},
  pages        = {1073--1082},
  year         = {1957},
  doi          = {10.2140/pjm.1957.7.1073},
}

@article{Girardin04,
  author       = {Girardin, Valerie},
  title        = {Entropy maximization for {M}arkov and semi-{M}arkov processes},
  journal      = {Methodology and Computing in Applied Probability},
  publisher    = {Springer},
  volume       = {6},
  pages        = {109--127},
  year         = {2004},
}

@article{Hall35,
  author       = {Hall, Philip},
  title        = {On Representatives of Subsets},
  journal      = {Journal of the London Mathematical Society},
  volume       = {s1-10},
  number       = {1},
  pages        = {26--30},
  year         = {1935},
  doi          = {10.1112/jlms/s1-10.37.26},
}

@article{Jaynes57,
  author       = {Jaynes, Edwin T.},
  title        = {Information theory and statistical mechanics},
  journal      = {Physical Review},
  publisher    = {APS},
  volume       = {106},
  number       = {4},
  pages        = {620--630},
  year         = {1957},
}

@article{Justesen84,
  author       = {Justesen, J{\o}rn and Hoholdt, T.},
  title        = {Maxentropic {M}arkov chains},
  journal      = {IEEE Transactions on Information Theory},
  publisher    = {IEEE},
  volume       = {30},
  number       = {4},
  pages        = {665--667},
  year         = {1984},
}

@article{Kalbfleisch83,
  author       = {Kalbfleisch, John David and Lawless, Jerald F. and Vollmer, William M.},
  title        = {Estimation in {M}arkov models from aggregate data},
  journal      = {Biometrics},
  publisher    = {JSTOR},
  pages        = {907--919},
  year         = {1983},
}

@book{Kemeny69,
  title={Finite Markov Chains},
  author={Kemeny, John G. and Snell, J. Laurie},
  year={1960},
  publisher={D. Van Nostrand Company},
  address={Princeton, NJ}
}

@inproceedings{Kumar15,
  author       = {Kumar, Ravi and Tomkins, Andrew and Vassilvitskii, Sergei and Vee, Erik},
  title        = {Inverting a steady-state},
  booktitle    = {Proceedings of the Eighth ACM International Conference on Web Search and Data Mining},
  pages        = {359--368},
  year         = {2015},
}

@book{Luce59,
  author       = {Luce, R. Duncan},
  title        = {Individual Choice Behavior: A Theoretical Analysis},
  publisher    = {John Wiley \& Sons},
  address      = {New York},
  year         = {1959},
}

@article{Mascherpa26,
  author       = {Mascherpa, Michele and Ringh, Axel and Taghvaei, Amirhossein and Karlsson, Johan},
  title        = {A convex approach for {M}arkov chain estimation from aggregate data via inverse optimal transport},
  journal      = {European Journal of Control},
  publisher    = {Elsevier},
  pages        = {101592},
  year         = {2026},
}

@article{Parry64,
  author       = {Parry, William},
  title        = {Intrinsic {M}arkov chains},
  journal      = {Transactions of the American Mathematical Society},
  publisher    = {JSTOR},
  volume       = {112},
  number       = {1},
  pages        = {55--66},
  year         = {1964},
}

@article{Pastor15,
  author       = {Pastor-Satorras, Romualdo and Castellano, Claudio and Van Mieghem, Piet and Vespignani, Alessandro},
  title        = {Epidemic processes in complex networks},
  journal      = {Reviews of Modern Physics},
  publisher    = {APS},
  volume       = {87},
  number       = {3},
  pages        = {925--979},
  year         = {2015},
}

@article{Proskurnikov17,
  author       = {Proskurnikov, Anton V. and Tempo, Roberto},
  title        = {A tutorial on modeling and analysis of dynamic social networks. {P}art {I}},
  journal      = {Annual Reviews in Control},
  publisher    = {Elsevier},
  volume       = {43},
  pages        = {65--79},
  year         = {2017},
}

@article{Shannon48,
  author       = {Shannon, Claude Elwood},
  title        = {A mathematical theory of communication},
  journal      = {Bell System Technical Journal},
  publisher    = {Nokia Bell Labs},
  volume       = {27},
  number       = {3},
  pages        = {379--423},
  year         = {1948},
}

@article{Sinkhorn67,
  author       = {Sinkhorn, Richard and Knopp, Paul},
  title        = {Concerning nonnegative matrices and doubly stochastic matrices},
  journal      = {Pacific Journal of Mathematics},
  publisher    = {Mathematical Sciences Publishers},
  volume       = {21},
  number       = {2},
  pages        = {343--348},
  year         = {1967},
}

\end{document}